\documentclass[preprint2]{aastex}
\usepackage[mathscr]{eucal}
\usepackage{graphics}

\lefthead{Guyon}
\righthead{Coronagraphy}

\begin{document}

\title{Theoretical Limits on Extrasolar Terrestrial Planet Detection with Coronagraphs}

\author{O. Guyon\altaffilmark{1}, E.A. Pluzhnik\altaffilmark{1}, M.J. Kuchner\altaffilmark{2}, B. Collins\altaffilmark{3}, S.T. Ridgway\altaffilmark{4}}
\altaffiltext{1}{Subaru Telescope, National Astronomical Observatory of Japan, 650 N. A'ohoku Pl., Hilo, HI 96720. USA.}
\altaffiltext{2}{NASA/Goddard Space Flight Center, Greenbelt, MD 20771, USA.}
\altaffiltext{3}{Research Institute for Mathematical Sciences, Kyoto University, Kyoto, 606-8502, JAPAN}
\altaffiltext{4}{National Optical Astronomical Observatory, 950 North Cherry Avenue, Tucson, AZ 85719, USA}

\begin{abstract}
Many high contrast coronagraph designs have recently been proposed. In this paper, their suitability for direct imaging of extrasolar terrestrial planets is reviewed. We also develop a linear-algebra based model of coronagraphy that can both explain the behavior of existing coronagraphs and quantify the coronagraphic performance limit imposed by fundamental physics. We find that the maximum theoretical throughput of a coronagraph is equal to one minus the non-aberrated non-coronagraphic PSF of the telescope. We describe how a coronagraph reaching this fundamental limit may be designed, and how much improvement over the best existing coronagraph design is still possible. Both the analytical model and numerical simulations of existing designs also show that this theoretical limit rapidly degrades as the source size is increased: the ``highest performance'' coronagraphs, those with the highest throughput and smallest Inner Working Angle (IWA), are the most sensitive to stellar angular diameter. This unfortunately rules out the possibility of using a small IWA ($<\lambda/d$) coronagraph for a terrestrial planet imaging mission. 

Finally, a detailed numerical simulation which accurately accounts for stellar angular size, zodiacal and exozodiacal light is used to quantify the efficiency of coronagraph designs for direct imaging of extrasolar terrestrial planets in a possible real observing program. We find that in the photon noise limited regime, a 4m telescope with a theoretically optimal coronagraph is able to detect Earth-like planets around 50 stars with 1hr exposure time per target (assuming 25\% throughput and exozodi levels similar to our solar system). We also show that at least 2 existing coronagraph design can approach this level of performance in the ideal monochromatic case considered in this study.
\end{abstract}


\section{Introduction}
Direct imaging and characterization (through low resolution spectroscopy) of extrasolar terrestrial planets (ETPs) is an exciting but challenging scientific application of coronagraphy. While the angular separation is well within the reach of a mid-sized telescope in the visible, the high star-planet contrast (about $10^{10}$ in the visible) requires a wavefront stability that can only be obtained with a space telescope. The low apparent luminosity of an Earth-like planet ($m_V \approx 29$ for an Earth at 10pc) requires the precious planetary photons to be well isolated from the stellar light; otherwise, photon noise from the stellar light drives the exposure time unreasonably high and speckles from the stellar image, if not calibrated, prevent detection. A system capable of delivering $10^{10}$ contrast at less than 0.1'' is therefore required for efficient detection and characterization of ETPs around a reasonably large ($\approx$30) sample of nearby stars.

The purpose of this work is to identify and quantify how fundamental physics imposes hard limits on what coronagraphs can achieve for the detection of terrestrial planets, and to compare these limits with what existing coronagraph designs can achieve.

Coronagraphs designs simulated in this work are introduced and \S\ref{sec:existingC}, where they are tentatively grouped in a four families. The configuration adopted for each design in this study is also described. In \S\ref{sec:idealcase}, a metric to quantify coronagraph performance is proposed, allowing existing coronagraph designs to be evaluated in the ``ideal'' case (monochromatic light, unresolved central star). A theoretical limit of coronagraph performance in this ideal case is also derived. In the next 2 sections, we closely look at the most fundamental (unavoidable) deviations from this ideal case, and their effects on coronagraphic performance: stellar angular size (\S\ref{sec:stellarsize}) and presence of a zodiacal/exozodiacal background (\S\ref{sec:zodi}). Finally, a detailed simulation of coronagraphic observations of a sample of nearby stars is performed in \S\ref{sec:montecarlo} to derive the efficiency of coronagraphs for detection of ETPs.

\section{Brief description of existing coronagraph designs}
\label{sec:existingC}
\begin{figure*}[htb]
\includegraphics[scale=0.55]{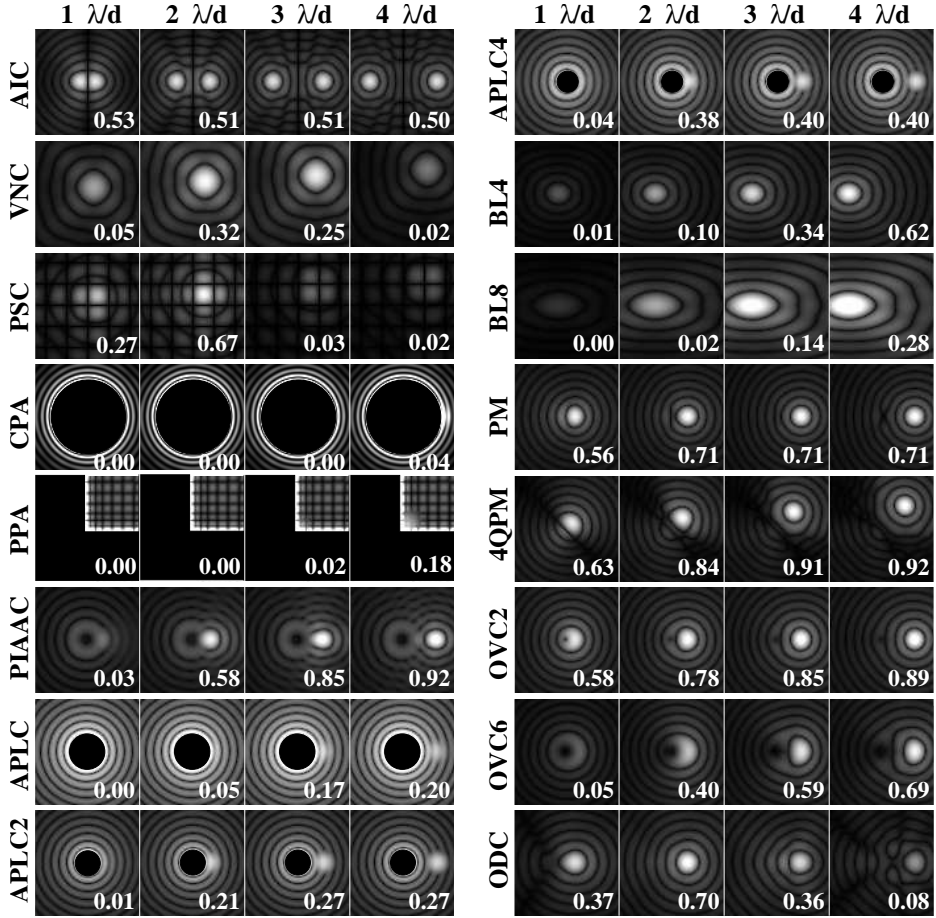}
\caption{\label{fig:psfs} Simulated monochromatic images of a $10^{10}$ contrast system, 1$\lambda/d$ to 4$\lambda/d$ separation, for some of the coronagraphs listed in Table \ref{tab:coronagraphs}. The central source was assumed to be a single point (angular diameter = 0). The white number in each image is the coronagraph throughput for the off-axis source. The pixel scale is the same for all images, but the brightness scale is not. The companion is moving on a diagonal (rather than horizontal) line for the VNC, PSC, PPA and 4QPM coronagraphs.}
\end{figure*}

The coronagraphs studied in this work (see Table \ref{tab:coronagraphs}) are all the coronagraphs known to us to theoretically achieve a $10^{10}$ PSF contrast within 5 $\lambda/d$ of the central source, as illustrated in Figure \ref{fig:psfs}. It is impossible to compare all the different kinds of coronagraph designs that have been suggested in all their details.  But without these concrete examples, we could not communicate our new perspective and illustrate what are the fundamental physical effects limiting coronagraphic performance.  We offer this brief summary of previous work as historical background prior to using these designs. More detailed descriptions of these coronagraphs can be found in the references given in this section.

We only consider in this paper unobstructed circular pupils. We note that several techniques have been proposed to adapt "unfriendly" pupil shapes (central obstruction, spiders) to specific coronagraphs \citep{soum05,mura05}.

\begin{deluxetable}{ l c c c }
\tablewidth{0pt}
\tabletypesize{\tiny}
\tablecaption{\label{tab:coronagraphs}Coronagraphs able to achieve $10^{10}$ PSF contrast within 5 $\lambda/d$}
\tablehead{ \colhead{Coronagraph} & \colhead{abrev.} & \colhead{reference} & \colhead{Design(s) adopted}}
\startdata
\multicolumn{4}{l}{{\bf ``Interferometric'' Coronagraphs}}\\
Achromatic Interferometric Coronagraph & AIC & \cite{baud00} & \\
Common-Path Achromatic Interferometer-Coronagraph & CPAIC & \cite{tavr05} & (=AIC)\\ 
Visible Nulling Coronagraph, X-Y shear ($4^{th}$ order null)\tablenotemark{a} & VNC & \cite{menn03} & Shear distance = $\pm0.3$ pupil radius\\
Pupil Swapping Coronagraph & PSC & \cite{guyo06} & Shear distance = 0.4 pupil diameter\\
\\
\multicolumn{4}{l}{{\bf Pupil apodization}}\\
Conventional Pupil Apodization and Shaped-Pupil\tablenotemark{b} & CPA & \cite{kasd03} & Prolate\tablenotemark{c} ($r=4.2\lambda/d$, 8\% throughput) \\
Achromatic Pupil Phase Apodization & PPA & \cite{yang04} & $\phi = \phi_2(x)+\phi_2(y)$; $a=2$; $\epsilon=0.01$\\
Phase Induced Amplitude Apodization Coronagraph & PIAAC & \cite{guyo03} & Prolate apodization\\
Phase Induced Zonal Zernike Apodization & PIZZA & \cite{mart04} & Not simulated\\
\\
\multicolumn{4}{l}{{\bf Improvement on the Lyot concept with amplitude masks}}\\
Apodized Pupil Lyot Coronagraph & APLC & \cite{soum03} & $r=1.8\lambda/d$ \\
Apodized Pupil Lyot Coronagraph, $N$ steps & APLC$_N$ & \cite{aime04} & $(N,r) = (2,1.4);(3,1.2);(4,1.0)$\\
Band limited, $4^{th}$ order\tablenotemark{a} & BL4 & \cite{kuch02} & $\sin^4$ intensity mask, $\epsilon = 0.21$\\
Band limited, $8^{th}$ order & BL8 & \cite{kuch05} & $m=1, l=3, \epsilon=0.6$\\
\\
\multicolumn{4}{l}{{\bf Improvement on the Lyot concept with phase masks}}\\
Phase Mask & PM & \cite{rodd97} & with mild prolate pupil apod.\\
4 quadrant & 4QPM & \cite{roua00} & \\
Achromatic Phase Knife Coronagraph & APKC & \cite{abe01} & (=4QPM)\\ 
Optical Vortex Coronagraph, topological charge $m$ & OVC$_m$ & \cite{pala05} & $m=2,4,6,8$\\
Angular Groove Phase Mask Coronagraph & AGPMC & \cite{mawe05} & (=OVC)\\
Optical Differentiation & ODC & \cite{oti05} & mask: $x\times \exp{^{-(x/10)^2}}$\tablenotemark{d}\\
\\
\enddata
\tablenotetext{a}{The Visible Nulling Coronagraph (VNC) and Band limited $4^{th}$ order (BL4) coronagraphs belong to the same class of pupil-shearing $4^{th}$ order coronagraphs, and are simply 2 ways of achieving the same result. They can be designed to have exactly the same performance. In this Table, the VNC is chosen with a small IWA and 2 orthogonal shear directions, while the BL4 is designed with a larger IWA and 2 shears in the same direction. To reflect this similarity, they are referred to as VNC/BL4(1) for the small IWA option (listed as VNC in this Table) and VNC/BL4(2) for the large IWA option (listed as BL4 in this Table).}
\tablenotetext{b}{The CPA design adopted here is a continuous apodization (rather than binary apodization/shaped pupil) which maximizes the radially averaged performance at $\approx 4 \lambda/d$. More optimal designs exist in other conditions: CPA with high contrast at specific position angles for observations at $\approx 3 \lambda/d$ or high throughput CPA for observations at $>4 \lambda/d$.}
\tablenotetext{c}{CPA, APLC, APLC$_N$: $r$ is the radius, in $\lambda/d$, of the mask within which the circular prolate function is invariant to a Hankel transform. This parameter is half of the mask diameter $a$ defined in \cite{soum03}.}
\tablenotetext{d}{ODC: x is in $\lambda/d$. Maximum mask transmission at $7 \lambda/d$. Lyot pupil mask radius = 0.85 times pupil radius.}
\end{deluxetable}

\subsection{``Interferometric'' coronagraphs (AIC, VNC, PSC)}
These coronagraphs look much like nulling interferometers: they rely on interferometric combination of discrete beams derived from the entrance pupil. 
\begin{itemize}
\item{AIC: The Achromatic Interferometric Coronagraph \citep{gay96,baud00} uses a beam splitter to destructively combine 2 copies of the entrance pupil, one of them achromatically $\pi$-phase shifted and flipped. The final image exhibits a 180 degree ambiguity which may be removed at the expense of loosing achromaticity \citep{baud05}. The Common Path Achromatic Interferometric Coronagraph (CPAIC) developed by \cite{tavr05} achieves the same achromatic $\pi$-phase shifted and flipped nulling with a common path interferometer, and is therefore optically more robust.}
\item{VNC: The Visible Nulling Coronagraph, $4^{th}$ order \citep{menn03} is the coronagraph equivalent of a double-Bracewell nulling interferometer. Two successive shears in perpendicular directions produce 4 beams, which, when combined, yield a $4^{th}$ order null in the pupil plane. By producing an image from the nulled pupil, this coronagraph combines a deep null with good imaging capabilities. In this paper, the shear distance is chosen to be 30\% of the pupil size, which places the first transmission maximum (36\% throughput) at $2.35 \lambda/d$ from the optical axis. Increasing the shear distance allows smaller IWA, but reduces the throughput. We note that another coronagraph, the $4^{th}$ order band limited coronagraph (see \S\ref{ssec:lyot}) could be designed to perform the same pupil shear and nulling operations. Both designs are simply two different optical implementations of the same principle, and are referred to as VNC/BL4(1) in this paper.}
\item{PSC: In the Pupil Swapping Coronagraph \citep{guyo06}, parts of the pupils are geometrically swapped prior to destructive interferometric combination, thus avoiding the throughput loss due to the shear in the VNC. The PSC design considered in this work is the $4^{th}$ order PSC described in \cite{guyo06}.}
\end{itemize}

\subsection{Pupil apodization coronagraphs (CPA, PPA, PIAA, PIZZA)}
The pupil complex amplitude can be modified to yield a PSF suitable for high contrast imaging, a property used by many coronagraph concepts.

\subsubsection{Conventional Pupil Apodization (CPA) with amplitude masks}
\label{sec:CPAbinarymask}
Apodization can be performed by a pupil plane amplitude mask (Conventional Pupil Apodization - CPA), which can be continuous \citep{jacq64,nise01,gons03,aime05b} or binary \citep{kasd03,vand03a,vand03b,vand04,aime05b}. Apodization by Mach-Zehnder type pupil plane interferometry was also suggested by \cite{aime01} to produce a continuous apodization, but is not explored in this work.

Within the CPA ``family'', we adopt for our comparative study a prolate spheroidal apodized pupil design \citep{slep65,aime02,kasd03,soum03,vand03b}, optimally tuned, according to the performance criteria defined in \S\ref{sec:usefulT}, for a $4 \lambda/d$ angular separation. In this design, the apodizer transmits 8\% of the light, and at $4 \lambda/d$ separation, exactly half of this light is detectable according the the criteria defined in \S\ref{sec:usefulT}. This apodization is nearly identical to the general prolate apodization function shown in Figure 8 of \cite{vand03a}.

\subsubsubsection{Design tradeoffs}
In CPAs, tradeoffs exist between IWA, throughput and ``discovery space'' (fraction of the field of view usable for planet detection). A description of these tradeoffs can be found in \cite{kasd03}, and they have been extensively explored for the design of binary masks/shaped pupils \citep{vand03a,vand03b,vand04}.
For example, in the checkerboard-mask design proposed in Figure 1 of \cite{vand04}, the PSF surface brightness reaches $10^{-10}$ at $2.8 \lambda/d$, but at this separation, the companion needs to be on the diagonal to be observed, and only a quarter of its light then falls in a ``black'' quadrant. In this particular small-IWA design, at 2.8 $\lambda/d$, the peak throughput is then 3.8\% for a square pupil, and the radially averaged throughput is $\approx$2\%. At large ($>4 \lambda/d$) separation, the average and peak throughputs are 15.1\%. 
Our prolate spheroidal apodization is not the optimal design if peak throughput is to be maximized (in this case, the ``single prolate'' shaped pupil described in \cite{kasd03} seems much superior). Our choice is therefore not optimal if the companion's position is previously known, in which case it could be intentionally placed at a given position angle.

\subsubsubsection{Binary vs. continuous apodization}
Binary pupil masks have the advantage of being achromatic and significantly easier to manufacture than continuous transmission masks. This technological advantage however comes at a non-negligible cost in planet detection performance:
\begin{itemize}
\item{In the PSF delivered by a non-obstructed, non-apodized circular aperture (Airy pattern), 84\% of the energy transmitted by the pupil is contained in the central diffraction spot; if the pupil is apodized with a continuous mask, this fraction is close to 100\% (but the spot is larger and less light is transmitted by the mask). In the PSF delivered by binary apodization masks, this fraction is typically about 50\%: there is a factor $\approx 2$ difference between the mask throughput and the ``Airy throughput'' \citep{kasd03} useful for detection. Within the central diffraction spot of a planet's image, the planet flux is proportional to the Airy throughput while the background light contribution is proportional to the mask throughput. Binary masks therefore ``amplify'' the amount of background light mixed with the planet image. This effect, discussed in \S\ref{sec:DEF} is also found in sparse aperture interferometers, where additional background light is collected from the secondary lobes.}
\item{The difference between mask throughput and Airy throughput is diffracted in regions of the PSF where planet detection is impossible. In a properly designed mask, this unwanted light avoids a large central region around the PSF's center.}
\end{itemize}
As the number of openings or slits in a binary mask increases, these two effects become less of a concern, and the mask asymptotically becomes identical to a continuous mask. 

\subsubsection{Pupil Phase Apodization (PPA)}
If only one half of the focal plane is considered, an amplitude pupil apodization can be replaced by a phase-only apodization (Pupil Phase Apodization, PPA), just as phase corrections in the pupil can only cancel focal plane speckles in one half of the field of view. High contrast imaging with phase apodization was proposed by \cite{yang04}, who found solutions for broadband imaging and obtained contrast/throughput performances similar to amplitude apodization designs (although only over a quarter of the field of view). \cite{codo04} independently computed a PPA solution for the Hubble Space Telescope pupil to suppress diffraction in half of the field of view. The PPA design adopted in this work uses a square pupil mask and the sum of two identical phase apodizations (equation 6 in \cite{yang04}), one in the x-direction, the other in the y-direction.

Although the PPA mask has a 100\% throughput within the square aperture, it exhibits large phase slopes at its edges, which are necessary to suppress diffraction in a quarter of the field but significantly increase the intensity of diffracted light in other parts of the field, reduce the amount of light in the central core of the PSF and broaden the PSF core. Just as for binary amplitude apodization masks, the PPA's large ratio between the pupil mask throughput (here equal to ~0.6 - area of the largest square that fits within the pupil) and the light concentrated in the PSF core ($\approx 0.3$) is problematic in the presence of zodiacal and exozodiacal backgrounds.

\subsubsection{Phase-Induced Amplitude Apodization Coronagraph (PIAAC)}
This coronagraph uses a lossless amplitude apodization of the pupil performed by geometric redistribution of the light rather than selective absorption \citep{guyo03,traub03,guyo05a,vand05,mart06,vand06,pluz06}. The same beam remapping technique is known as beam shaping in laser science \citep{shea02} and radio astronomy.

The geometrical remapping of the telescope pupil amplifies phase slopes, therefore enabling observations at angular separations $\approx$ 3 times smaller than with a conventional amplitude apodization. 
Two designs are considered in this work:
\begin{itemize}
\item{{\bf PIAA:} The detection occurs in the first focal plane after the beam remapping. Starlight therefore falls on the detector but is concentrated in a single diffraction spot. This design is adopted for practical reasons in \S\ref{sec:idealcase} and \S\ref{sec:stellarsize} to avoid having to tune a focal plane mask diameter.}
\item{{\bf PIAAC:} A focal plane mask occults starlight in the first focal plane. An inverse beam remapping unit then restores the original pupil to recover field of view. While a focal plane mask size needs to be chosen for this design, it is adopted in \S\ref{sec:montecarlo} to maintain a sharp planet image over the zodi and exozodi backgrounds.}
\end{itemize}

\subsubsection{PIZZA}
An interesting alternative to CPA techniques is Phase Induced Zonal Zernike Apodization, PIZZA, \citep{mart04}, which achieves the pupil amplitude apodization without compromising the telescope's throughput or angular resolution. PIZZA, which is not simulated in this work, may be too chromatic to be practically suitable for imaging ETPs. The IWA delivered by PIZZA, CPA and PPA are similar.

\subsection{Improvements of the Lyot coronagraph design (APLCs, BLs, PM, 4QPM, OVCs, ODC)}
\label{ssec:lyot}
In a Lyot coronagraph, the telescope pupil and the focal plane mask are not perfectly fitted one for another, resulting in strong stellar leaks (much higher than the $10^{-10}$ goal) within the coronagraph's Lyot stop. For a Lyot coronagraph to deliver high contrast performance, the complex amplitude in the ``intermediate'' focal plane (equal to the Fourier transform of the entrance pupil multiplied by the focal plane occulter) needs to be ``band-limited'': its Fourier transform ($=$ complex amplitude in the Lyot pupil plane) should have nearly zero power in a frequency range which is then selectively transmitted by the Lyot pupil stop. Improved performance over the original Lyot design can therefore be obtained by either :
\begin{itemize}
\item{Adapting the pupil to the hard edge focal plane mask. This is achieved in the Apodized Pupil Lyot Coronagraph, APLC, described in \cite{soum03}, where the entrance pupil of a Lyot coronagraph with a hard edge focal plane occulter is optimally apodized. A slightly different approach, explored by \cite{vand04} (see Figure 3 of the paper), is to apodize the pupil after the hard edge focal plane occulter. As suggested by \cite{aime04}, the output of an APLC can be used as the input of a second stage APLC: these are the multistep APLCs (noted APLC$_N$, where $N$ is the number of steps).}
\item{Adapting the focal plane mask to the telescope pupil. This is achieved by using ``band limited masks'' in the focal plane: the $4^{th}$ order band limited coronagraph, BL4 \citep{kuch02} and 8th order band limited coronagraph, BL8 \citep{kuch05}. A visible nuller interferometer (VNC) equivalent to the BL4 coronagraph design adopted in this work could be designed, and we therefore refer to this BL4 design as VNC/BL4(2).}
\end{itemize}

Phase mask coronagraphs (PM, 4QPM, OVC, ODC) introduce phase shifts in the focal plane. This allows smaller IWAs than offered by APLCs and BLs.\begin{itemize}
\item{PM: The Phase Mask coronagraph \citep{rodd97} uses a circular $\pi$-shifting focal plane mask, and, to allow $10^{10}$ contrast, a mild pupil amplitude apodization \citep{guyo00,soum03}.}
\item{4QPM: The 4-Quadrant Phase Mask \citep{roua00} uses a focal plane mask which shifts 2 out of 4 quadrants of the image by $\pi$. The achromatic phase knife coronagraph \citep{abe01} is another form of 4QPM.}
\item{OVC: In the Optical Vortex Coronagraph \citep{pala05,foo05,swar06} and the Angular Groove Phase Mask Coronagraph \citep{mawe05}, a focal plane vortex phase mask replaces the 4 quadrant phase mask of the 4QPM, thus avoiding the ``dead zones'' of the 4QPM. In $(r,\theta)$ polar coordinates, the mask phase is equal to $m\theta$, where $m$ is the {\it topological charge}. OVCs with high topological charges ($m>2$) exhibit low sensitivity to low-order aberrations.}
\item{ODC: The Optical Differentiation Coronagraph \citep{oti05} is using a phase and amplitude focal plane mask.}
\end{itemize}

\subsection{Other coronagraph designs \& techniques not studied in this work}
Some of the designs listed in Table \ref{tab:coronagraphs} may be combined in series. The multistep APLC concept \citep{aime04} illustrates how such a ``cascade'' can simultaneously improve IWA and throughput by designing each step for a lower contrast level. \cite{nish05} showed how a nulling interferometer can benefit from a CPA, and a similar combination may also be advantageous on a circular pupil (VNC-CPA for example). The entrance pupil apodization required for the APLCs may be performed by a PIAA unit to improve IWA and throughput.
Although combinations of coronagraphic techniques are not studied in this paper, the results obtained on individual designs can provide a good insight into which combinations may be most beneficial.

Pupil replication \citep{gree05} has been proposed to transform a phase slope (equivalent to angular separation on the sky) into phase discontinuities in the replicated pupil. For the off-axis planet, this creates secondary diffraction peaks far from the central source image. The authors pointed out that it could enhance the performance of a coronagraph with which it would be combined. \cite{riau05}, in a sensitivity study of pupil replication, suggested to use it after a 4QPM. The tip-tilt sensitivity of both techniques (pupil replication introduces phase steps in the wavefront of an off-axis source; see \S\ref{sec:ttsens} for the 4QPM) may be a concern for direct imaging of terrestrial planets.

A coronagraphic effect can be obtained by placing an occulting screen, at least as big as the telescope diameter, between the star and the telescope \citep{cash05}. This external occulter needs to be designed to not diffract starlight within the telescope aperture, needs to be placed far from the telescope (on a separate spacecraft), and the telescope-occulter-star alignment needs to be maintained during observations. While maintaining the alignment is especially challenging, a ``conventional'' imaging telescope can be used without the need for exquisite control of wavefront aberrations.

\section{Coronagraph performance for a monochromatic point source}
\label{sec:idealcase}

\subsection{A coronagraph performance metric: the ''useful throughput''}
\label{sec:usefulT}

Since we aim at quantitatively comparing the performance of coronagraph designs, we need a ''fair'' performance metric: one that can be applied uniformly to all coronagraph designs. Coronagraph performance is usually quantified with Inner Working Angle (usually defined as the minimal angular distance at which the throughput for the planet is half of the maximal throughput), throughput (measured at large angular separation), and search area (fraction of the focal plane image within which a planet can be detected). The IWA is somewhat arbitrary, as there is no well-defined stellar-to-companion light ratio threshold beyond which companion light suddenly becomes unmeasurable. IWA is coupled with throughput: for example, a coronagraph with a 10\% maximum throughput achieving 5\% throughput at 1 $\lambda/d$ has a smaller IWA than a coronagraph with a 100\% maximum throughput achieving 30 \% throughput at 1 $\lambda/d$. In this example, the larger IWA coronagraph outperforms the smaller IWA coronagraph. Search area is also somewhat arbitrary and often cannot be quantified as a single number. For example, diffraction from 4 spider vanes in 2 perpendicular directions (we assume here that the width of the diffraction spikes created by the spiders is equal to the IWA of the coronagraph) covers $\approx 100\%$ of the field of view at the IWA and $\approx 0\%$ at large angular separations. {\bf Metrics commonly used to quantify coronagraph performances (IWA, throughput and search area) are therefore difficult to use to directly compare coronagraph designs.}

A simpler, more direct measure of how well a coronagraph can separate planet light from star light, the ``useful throughput'', is proposed here. It is defined as the amount of planet light which can be used toward detection, and is expressed as a function of the planet position relative to the star and the planet/star contrast. When averaged over all position angles, it conveniently quantifies the coronagraph performance with a single number (making it possible to directly compare coronagraph performances), function of the angular separation. This function accurately includes the effects of IWA, throughput and search area, and is a good estimation of the coronagraph performance. 

To compute this quantity, we must first define when planet light in the coronagraph detector plane becomes ``unusable'' because of excessive starlight mixed with it. Theoretically, direct detection of companions embedded in much brighter starlight is possible if the starlight component on the image is well characterized. Practically, due to calibration problems and photon noise (exoplanets are faint), companion light mixed with much brighter starlight is unusable. There is however no well-defined stellar-to-companion light ratio threshold beyond which companion light becomes unmeasurable: the criteria we will use to define the ''useful throughput'' (hereafter denoted throughput) is therefore somewhat arbitrary.

\begin{figure}[htb]
\includegraphics[scale=0.33]{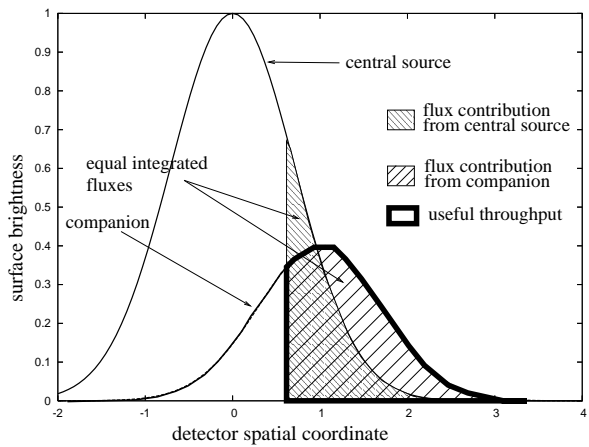}
\caption{\label{fig:usablet} Graphical representation of the useful throughput. In this 1-D example, the stellar and planet PSF are shown with some overlapping. The useful throughput is obtained by integrating the companion (planet) light from $x \approx 0.7$ to $x \approx 3.2$; in this interval, the integrated flux contributions from the central source and the companion are equal.}
\end{figure}

As illustrated in Figure \ref{fig:usablet}, the ``useful throughput'' is the maximum fraction of the planet's light gathered by the telescope which can be separated from starlight. The ``separation'' criteria used is that, integrated over the area of the focal plane considered, there is as much planet light as there is residual stellar light. Physically, this area may be a single pixel or a group of pixels on a focal plane array detector. 

To compute the useful throughput of a coronagraph at a particular location in the image plane, we first model the images $I_s(x,y)$ and $I_p(x,y)$ produced respectively by a star and a much fainter planet in the coronagraph. Pixels are then sorted in decreasing order of $I_p(x,y)/I_s(x,y)$; for the first pixels in this sorted list, a large fraction of the light collected originates from the planet while the last pixels of this list are dominated by starlight. The number $n$ is chosen such that, over the first $n$ pixels of this list, the total starlight collected is equal to the total planet light collected. This total amount of planet light, divided by the the flux in planet light that is collected by the entrance aperture, is the {\bf useful throughput}. We note that if, on every pixel of the image, starlight exceeds planet light, then the useful throughput will be zero.


\subsection{Useful throughput of existing coronagraph designs}

\begin{figure*}[htb]
\includegraphics[scale=0.33]{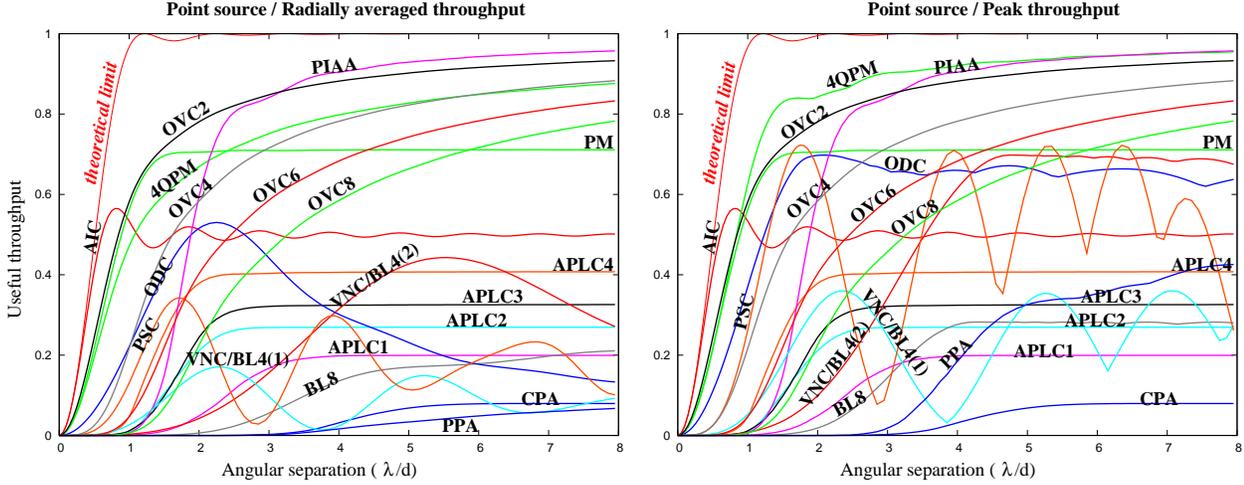}
\caption{\label{fig:usefulT} Throughput, at the $10^{10}$ contrast level, of the coronagraphs listed in Table \ref{tab:coronagraphs} as a function of angular separation. On the left, the useful throughput has been radially averaged for coronagraphs with ``preferential'' directions (BL4, BL8, 4QPM, ODC, VNC, PSC). On the right, the peak throughput is shown, assuming that the telescope orientation is optimal. The theoretical limit derived in \S\ref{sec:theoreticallimit_noT} is shown in red. The central source is assumed here to be monochromatic and infinitively small.}
\end{figure*}

Figure \ref{fig:usefulT} shows the useful throughput for the coronagraphs listed in Table \ref{tab:coronagraphs}. For small angular separations ($<1\lambda/d$) the AIC is the most efficient coronagraph, with a 50\% useful throughput at $0.5 \lambda/d$. Based on our assumptions, the phase mask coronagraphs (especially OVC$_2$, PM and 4QPM) and PIAA are the best choices from $0.5 \lambda/d$ outward. Many coronagraphs in this list never achieve a very high throughput because of strong entrance pupil apodization (CPA, APLC, and to a lesser degree APLC$_N$), beam splitting (PSC,VNC), or pupil diaphragming (VNC/BL4, BL8). The range of performance is very large: the AIC offers an IWA about 8 times smaller than the CPA, and the PIAA's throughput is about 10 times the CPA's throughput beyond $5\lambda/d$. 

Many of the coronagraph designs studied in this paper were developed in the last few years, and there is therefore hope that a large number of other coronagraph designs will be discovered in the near future. It is particularly interesting to wonder how much of the upper left area of Figure \ref{fig:usefulT} (high throughput at small angular distance) can/will be accessed.
In the following section, we derive a fundamental limit of coronagraph throughput (shown in Figure \ref{fig:usefulT}) using a ''universal'' model of coronagraphs.

\subsection{Theoretical upper limit for the coronagraph throughput}
\label{sec:theoreticallimit_noT}

\subsubsection{Coronagraph model}
\begin{figure*}
\includegraphics[scale=0.47]{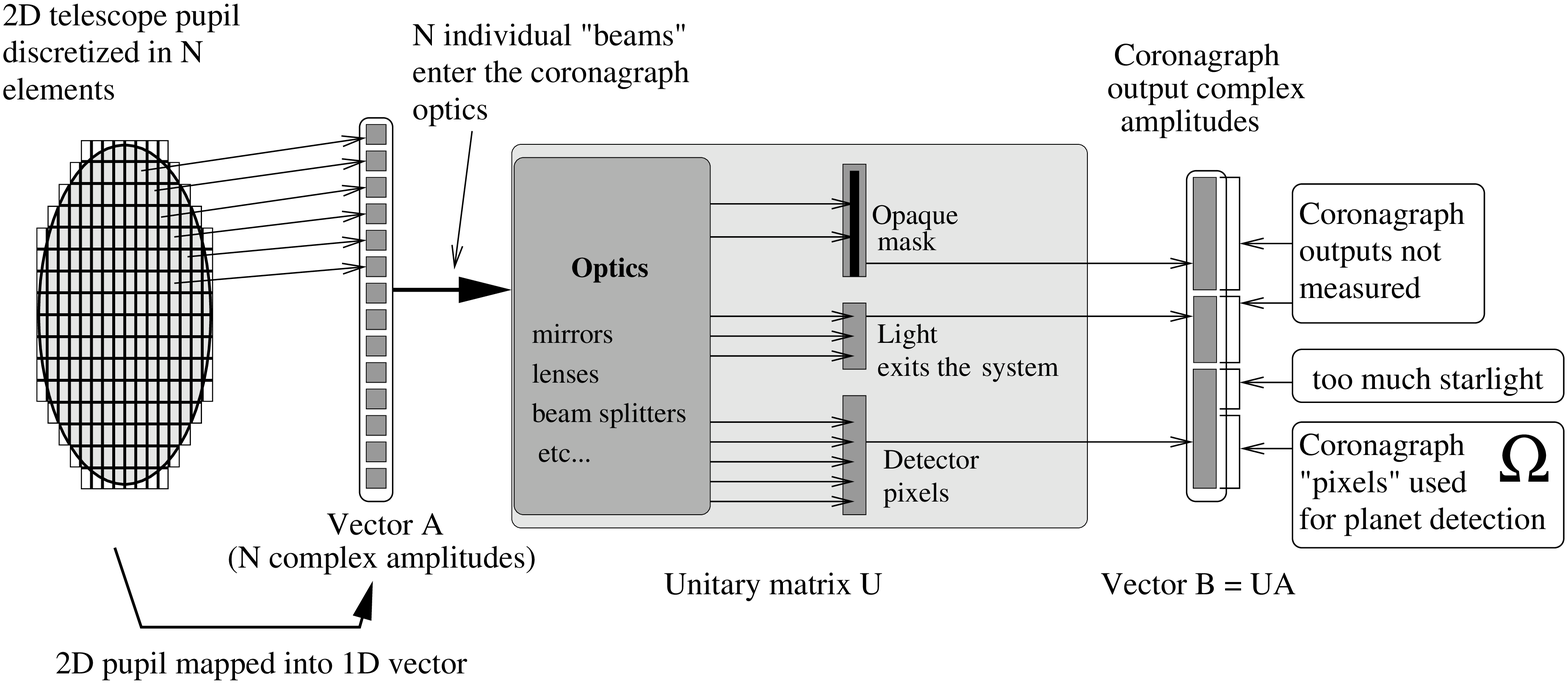}
\caption{\label{fig:matrixU} Algebraic representation of a coronagraph. As detailed in the text, the coronagraph can be represented by a complex unitary matrix $U$ which, when multiplied by the entrance complex amplitude vector $A$, yields the output complex amplitudes $B$. The goal of the coronagraph is to efficiently isolate starlight is some coefficients of $B$ while preserving as much of the companion's light is the other coefficients of $B$. The planet light useful for detection defines a subset $\Omega$ of coefficients in vector $B$.}
\end{figure*}

We consider an optical system, shown in Figure \ref{fig:matrixU}, in which light enters through a pupil (which may or may not be circular) which we represent by a finite (but large) number $N$ of regularly spaced elements. Similarly, the output of this optical system, defined by the points at which light is detected, blocked or exits the system, is represented by $N$ points.
To a coherent light source (such as a point source at infinity) corresponds a distribution of complex amplitude at the pupil of this system, written here as a vector $A$ of $N$ complex numbers. Although $A$ is written as a vector, it represents a 2D distribution of complex amplitude across the pupil: each of its coefficient is the complex amplitude of the incident wave across one of the $N$ elements in the entrance pupil. We note that this description is only valid to the extent that complex amplitude is constant across each pupil element - this limitation is equivalent to a field of view limit in the focal plane, and our model is therefore not suitable for point sources at more than $\approx \sqrt{N}\lambda/(2d)$ from the optical axis. For convenience, we choose to normalize this vector such that the total light intensity in the pupil is unity ($||A||^2 = 1$). 

Optical systems are (usually) linear in complex amplitude: the complex amplitude anywhere in the system can be written as a linear combination of complex amplitudes in the entrance pupil. For example, Fourier transforms and Fresnel propagation simulations are linear operations commonly performed to compute coronagraphic images.
The output complex amplitude vector $B$ can therefore be written as:
\begin{equation}
B(\alpha) = U A(\alpha)
\end{equation}
where $U$ is a NxN complex matrix which entirely describes the optical system and $\alpha$ is the point source position in the sky. The output vector $B$ includes light that is blocked by masks (turned into heat), exits the system (for example, reflected back to the sky) or falls on the detector (these are the only coronagraph outputs that can be measured and therefore used for planet detection). Conservation of energy in the system imposes $||B||^2 = ||A||^2$: $U$ is therefore a {\bf complex unitary matrix, function of the coronagraph optics, but independent of the entrance wavefront} (and therefore independent of the source position $\alpha$ in the sky). Wavefront control elements, if present, are also part of $U$.

As illustrated in Figure \ref{fig:matrixU}, in a coronagraph, the goal is to create system output(s) for which an on-axis source will produce very little light, but a closeby off-axis source's flux will be transmitted. Mathematically speaking, a matrix $U$ needs to be chosen such that some coefficients of $B(\alpha)$ (we denote $n<N$ the number of such coefficients) are very small for $\alpha \approx 0$, but relatively high for other values of $\alpha$. To this set of coefficients corresponds a sub-space $\Omega$ of dimension $n$ (shown as the bottom part of vector $B$ on Figure \ref{fig:matrixU}).

The coronagraph optimization problem can be further restricted by fixing a 2D vector $A(\alpha_p)$, where $\alpha_p$ is the planet off-axis position at which the coronagraph's throughput is to be maximized. In this case, a change of coordinate system (rotation) within the sub-space $\Omega$ can concentrate all the planet flux within $\Omega$ on a single coefficient. This change of coordinate system is a function of $\alpha_p$, and aligns, within $\Omega$, the 2D vector $A(\alpha_p)$ with one of the directions of the new coordinate system. Physically this is equivalent to placing multiple beam splitters and phase shifters to concentrate all of the planet light within $\Omega$ in a single output beam: this would be possible thanks to the fact that planet light is fully coherent.

This norm-preserving coordinate change is itself a unitary matrix which can be integrated within $U$ (by direct multiplication with the ``old '' $U$), in which case only a single column of this new $U$ matrix is now relevant to the problem: we denote $C$ this column ($C$ is a vector). Since $U$ is unitary, $||C|| = 1$. In this new basis, $|C \bullet A(\alpha)|^2$ is the square absolute value of one of the coefficients of the new vector $B$ and is equal to the coronagraph throughput at the sky position $\alpha$: it is therefore the quantity to be minimized for $\alpha = 0$ and maximized for $\alpha = \alpha_p$.

\subsubsection{Coronagraphic throughput upper limit in the ideal case.} 

We use in this section the new base described at the end of the previous section, where the sub-space $\Omega$ corresponds to a single coefficient of $B$. In this ``optimized coronagraphic projection'', the energy in this single coefficient is the useful throughput of the coronagraph, as long as, within this single ``pixel'', the stellar flux is smaller than the planet flux. Excluding all other coefficients of $B$, we can now equivalently refer to the coronagraph throughput or useful throughput.

We denote $\epsilon$ the on-axis coronagraph throughput (no more than $10^{-10}$ for a system designed to image ETPs in the visible) and $A(0)$ the input complex amplitude for an on-axis source. We note that $\epsilon$ is a throughput, not a contrast, and should therefore ideally be no more than $10^{-10}$ times the coronagraph throughput for the planet. It is now possible to design an ideal theoretical coronagraph for detecting a source at the position $\alpha_p$ by choosing a complex vector $C$ of norm 1 such that:
\begin{enumerate}
\item{$|C \bullet A(0)| / |C \bullet A(\alpha_p)| < \sqrt{\epsilon}$: this is the coronagraphic ``contrast'' at position $\alpha_p$.}
\item{$|C \bullet A(\alpha_p)|$ is as large as possible: this is the the square root of the coronagraph intensity throughput at position $\alpha_p$.}
\end{enumerate}

\begin{figure}[htb]
\includegraphics[scale=0.33]{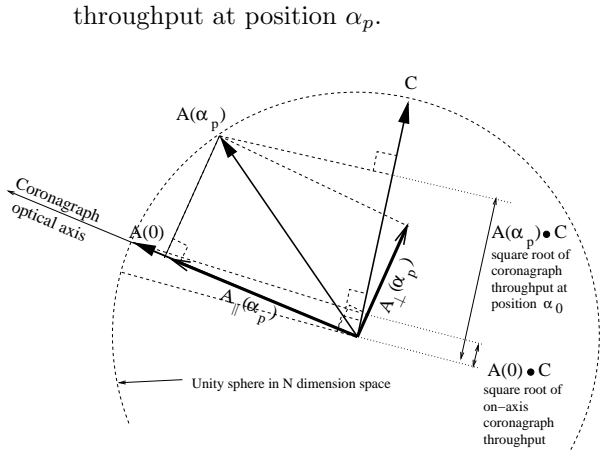}
\caption{\label{fig:projection} Graphical representation of the coronagraph optimization problem.}
\end{figure}

This way of looking at the problem shows that optimizing a coronagraph is equivalent to choosing a projection axis $C$ which is ``mostly'' perpendicular to $A(0)$ (constraint (1)) and as co-linear as possible with $A(\alpha_p)$ (constraint (2)). A glance at Figure \ref{fig:projection} reveals that if $A(\alpha_p)$ and $A(0)$ are close (ie. the planet position $\alpha_p$ is close to the optical axis), the coronagraphic throughput cannot be high.
As shown in Figure \ref{fig:projection}, the input vector $A(\alpha_p)$ can be decomposed into an ``along optical axis'' component and a ``perpendicular to optical axis'' component:
\begin{equation}
A(\alpha_p) = A_{\|}(\alpha_p) + A_{\perp}(\alpha_p)
\end{equation}
where 
\begin{equation}
\label{equ:A0a0}
A_{\|}(\alpha_p) = [A(\alpha_p) \bullet A(0)]\:A(0)
\end{equation}
is the projection of $A(\alpha_p)$ onto $A(0)$ (``along optical axis'').

Using this decomposition, the linearity of the coronagraph in complex amplitude, and the $|C \bullet A(0)|<\sqrt{\epsilon}$ constraint, an upper bound can be established for the square root of the coronagraphic throughput:
\begin{equation}
A(\alpha_p) \bullet C < [A_{\|}(\alpha_p) \bullet A(0)] \: \sqrt{\epsilon} + A_{\perp}(\alpha_p) \bullet C.
\end{equation}
Since $\epsilon << 1$ and $||C||=1$,
\begin{equation}
\label{equ:upplim}
[A(\alpha_p) \bullet C] < ||A_{\perp}(\alpha_p)||.
\end{equation}

\begin{figure}[htb]
\includegraphics[scale=0.3]{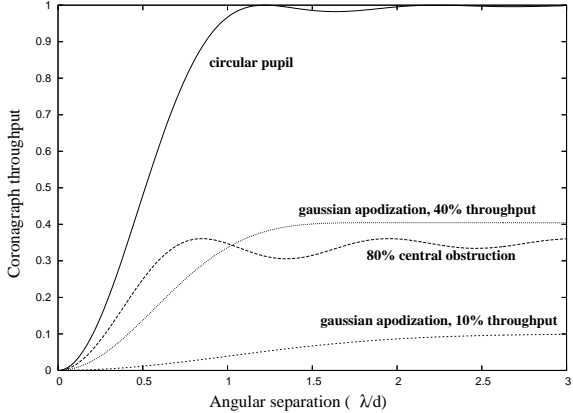}
\caption{\label{fig:upplim} Upper limit on the off-axis throughput of a coronagraph for different entrance pupils.}
\end{figure}

The upper limit given by equation \ref{equ:upplim} is shown in Figure \ref{fig:upplim} for a ``normal'' circular pupil (unobstructed, unapodized) as well as for Gaussian-apodized pupils and a pupil with a large 80\% central obstruction. Not surprisingly, amplitude apodization decreases the throughput at all distances and also pushes further the point at which the maximum throughput is reached (decrease in angular resolution). With a large central obstruction, the throughput reaches its maximum value quicker (slight increase in angular resolution).

This result shows that a coronagraph, regardless of its on-axis throughput (as long as it is much smaller than 1 of course), cannot have a throughput exceeding 50\% at 0.5 $\lambda/d$ for a circular pupil. Curiously, the upper limit derived by equation \ref{equ:upplim} oscillates with distance, especially in the large central obstruction case. This is due to a slight similarity in the wavefront at 0 $\lambda/d$ and $\approx 1.6 \lambda/d$ for the circular pupil case. In the extreme case of a 2-pupil interferometer of baseline $b$, the same effect would lead to a $\lambda/b$ periodicity of the theoretical maximum throughput.

Since $||A_{\perp}||^2 = ||A(\alpha_p)||^2-||A_{\|}||^2$ (right angle in Figure \ref{fig:projection}), $||A(\alpha_p)||=1$ by definition, and $||A_{\|}(\alpha_p)||^2$ is the intensity of the non-aberrated non-coronagraphic PSF at position $\alpha_p$ (equation \ref{equ:A0a0}):
{\bf the maximum throughput of a coronagraph is equal to one minus the non-aberrated non-coronagraphic PSF of the telescope.} In the case of a non-obstructed circular pupil, this throughput is one minus the Airy pattern, as can be seen in Figure \ref{fig:upplim}.

This last property is a direct consequence of the linearity in complex amplitude of the coronagraph. Since the coronagraph needs to remove the central star, it also removes the ``flat non-tilted component'' of any incident wavefront, which, expressed as a function of the source position on the sky, is the Airy function (for an unobstructed pupil). 

\subsubsection{Can this upper limit be reached in an optical system ?}
\label{ssec:canitbebuilt}
With the above analysis, the vector $C$ that would be required to reach the upper limit at a given position $\alpha_p$ can be computed. Each of the $N$ elements of the pupil can be treated as a small independent beam, and we could construct the optical system with a series of beam splitters (not necessarily 50\%/50\%) and phase shifts. Each beamsplitter combines 2 beams and produces 2 output beams. Since $[C \bullet C] = 1$, the problem is equivalent to concentrating all the flux into a single beam if the entrance vector ($N$ beams) is equal to $C$. It is relatively easy to do incrementally: at each step 2 beams are combined with a phase shift + beam splitter such that all the light is sent into one output beam (this is possible since the beams are fully coherent between them and we know exactly their amplitudes and phases). $N-1$ beam splitters are required in this scheme. If the coronagraph is only intended to work close to the optical axis, $N$ could be quite small ($N$ scales as the square of the outer working angle), resulting in an optical system that could possibly be built. Unfortunately, this coronagraph is only guaranteed to reach the theoretical limit at a single planet position $\alpha_p$.

A better way to build the ``ideal'' coronagraph is to have the light of an on-axis point source entirely concentrated in a single output. This would ensure that, for large values of $N$, the coronagraphic throughput (now measured on $N-1$ beams) is equal to the theoretical limit simultaneously at all source positions. Special attention would be given to the arrangement of the beam splitters that only affect the $N-1$ ``useful'' outputs in order to allow the source position to be easily retrieved from comparisons between their intensities.

\section{Sensitivity to tip-tilt: the stellar angular diameter problem}
\label{sec:stellarsize}

\subsection{Effect of stellar angular diameter in existing coronagraph designs}
\label{sec:ttsens}
The sensitivity of some coronagraphs to tip-tilt errors has been investigated by several authors (see for example \cite{lloy05,shak05,siva06}). 

A Sun-like star at 10pc is 1mas across: with an 8m visible telescope its apparent radius is about $0.1 \lambda/d$. Many coronagraphs cannot maintain a $10^{10}$ contrast on such an extended central source. Even if the telescope diameter were reduced (or the wavelength increased) such that the planet, at maximum elongation, would be at $2 \lambda/d$, the stellar radius would still be $0.01 \lambda/d$.

\begin{figure*}
\includegraphics[scale=0.33]{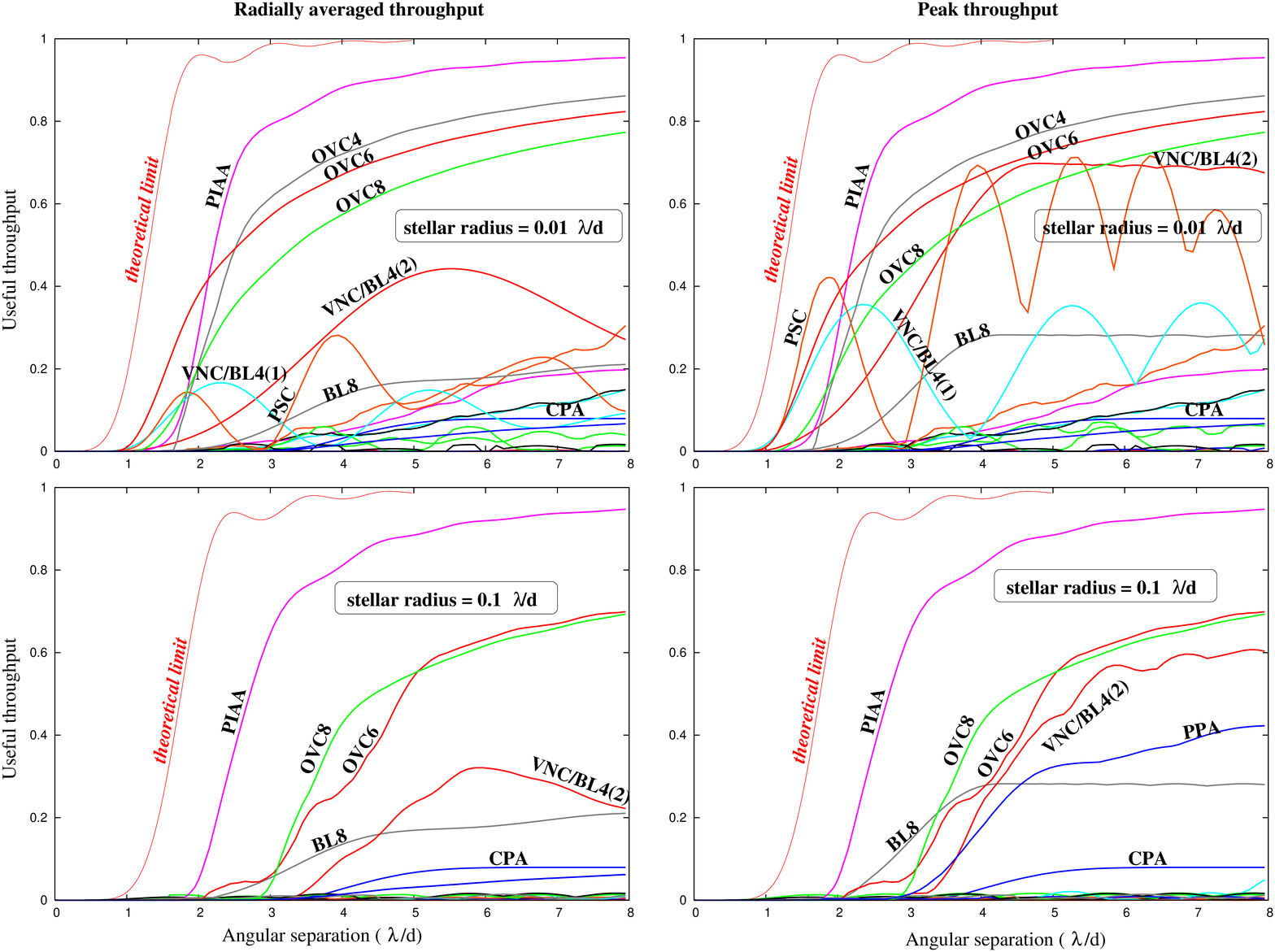}
\caption{\label{fig:usefulTtt} Same as Figure \ref{fig:usefulT}, but with an extended central source (top: $0.01 \lambda/d$ radius disk, bottom: $0.1 \lambda/d$ radius disk).}
\end{figure*}

Figure \ref{fig:usefulTtt} shows how coronagraphic performance (useful throughput) decreases as the stellar radius increases, and can be used to somewhat arbitrarily divide coronagraphs in 3 groups:
\begin{itemize}
\item{High sensitivity to stellar radius (AIC, 4QPM, PM, OVC$_2$, ODC). These coronagraphs cannot maintain high contrast even with a 0.01 $\lambda/d$ stellar radius. They would be very difficult to use for ETP imaging.}
\item{Moderate sensitivity to stellar radius (VNC and BL4 with small IWA/large shear: VNC/BL4(1), PSC, APLCs, OVC$_N$ with large value of $N$). These coronagraphs may be used efficiently on ``small'' diameter telescopes (for which the stellar radius is about $0.01 \lambda/d$ or less).}
\item{Low sensitivity to stellar radius (PIAAC, VNC and BL4 with large IWA/small shear, BL8, CPA). These coronagraphs are suitable for direct imaging and characterization of ETPs, as they tolerate stellar radii of $0.1 \lambda/d$.}
\end{itemize} 
The robustness of VNC/BL4(1) compared to the VNC/BL4(2) is entirely due to the larger IWA chosen in this this study for the VNC/BL4(2). For both the BL4 and VNC, there is a IWA vs. sensitivity to stellar radius tradeoff.

For most coronagraphs, increasing the stellar diameter first compromises contrast at small angular separations. Even the VNC/BL4(2) and PIAAC, which are relatively insensitive to angular size, lose throughput at small angular separations in the $0.1 \lambda/d$ radius case. 

The CPA and BL8 stand out as being extremely robust to stellar angular size: their throughput curves are almost identical from 0 to 0.1 $\lambda/d$ stellar radius. The CPA owes its relative immunity to tip-tilt errors to the field-invariance of its PSF. The sensitivity of the BL8 to tip-tilt errors was studied in detail by \cite{shak05}, who also find that the BL8 can tolerate relatively high levels of focus, coma, and astigmatism. 

Direct comparison between Figures \ref{fig:usefulT} and \ref{fig:usefulTtt} shows a very disappointing result: the the best coronagraphs in the point source case (AIC, PM, 4QPM, OVC2) all perform very poorly when the stellar angular size is considered. Increasing the stellar size from 0 to $0.01 \lambda/d$ has moved the smallest angular separation at which 50\% throughput is reached from 0.5 $\lambda/d$ (for the AIC) to slightly more than $2 \lambda/d$ (for the PIAA). Is this behavior imposed by fundamental physics and therefore unavoidable ? Or is there hope to find a low-IWA coronagraph which is not so dramatically sensitive to stellar angular size ?

\subsection{Throughput limit imposed by stellar angular size}
\label{sec:ttideal}
\begin{figure}[htb]
\includegraphics[scale=0.3]{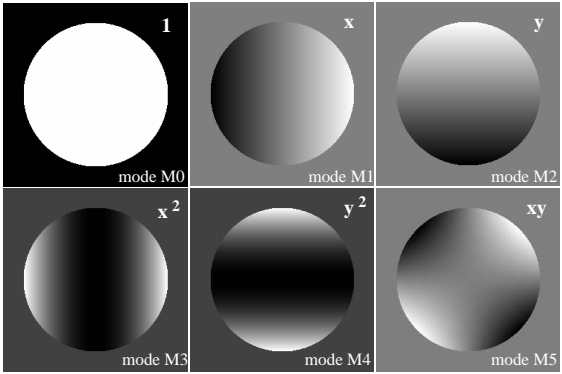}
\caption{\label{fig:modes} Graphical representation of the first 5 modes of the orthonormal base $M_i$ for a circular unobstructed pupil. The vectors $M_i$ are independent of stellar size, and their coefficients are all real numbers (no imaginary part).}
\end{figure}

In this section, we quantify how the angular extent of the source affects the coronagraph throughput. We assume that the coronagraph is observing a disk of uniform surface brightness and angular radius $r_s$. Using the same framework as in \S\ref{sec:theoreticallimit_noT}, the star is now modeled by a series of vectors $A_k = A(\alpha_k)$, where the $\alpha_k$ are uniformly distributed on the stellar disk ($|\alpha_k|<r_s$) and $k = 0 ... N_s-1$. An orthonormal base can be built from the vectors $A_k$ with the following process:
\begin{enumerate}
\item{The vector $M_i$ ($i=0$ initially) of this base is chosen to be co-linear with the vector $A_k$ with the highest norm.}
\item{All vectors $A_k$ are replaced by their projection on a hyperplane perpendicular to $M_i$.}
\item{$i$ is incremented and we return to step 1 with the new vectors $A_k$.}
\end{enumerate}
This algorithm insures that, in the orthonormal base obtained, the coefficients of the vectors $A_k$ decrease very rapidly. We denote $m_i$, $i=0 ... N-1$ the maximum absolute value of the coefficient $i$ in this new base across all $A_k$. The 2D representation of the first 5 vectors $M_i$ are shown in Figure \ref{fig:modes}.
The same modes can be obtained by linear expansion of the pupil complex amplitude $e^{i(xu+yv)}$ for small values of $u$ and $v$ (here, pupil plane coordinates are ($x$,$y$) and point source angular coordinates on the sky are ($u$,$v$)).

As previously shown, a coronagraph which needs to cancel the light across the stellar disk can be represented by a vector $C$ of norm 1. We denote $c_i$, $i=0...N-1$ the coefficients of $C$ in the new base of modes $M_i$. To insure that the coronagraph cancels the light of the star by a factor $\epsilon$, the following constraint on each $c_i$ can be imposed :
\begin{equation}
\label{equ:ciconst}
|c_i| < \sqrt{\epsilon} / m_i.
\end{equation}
This constraint is not rigorously equivalent to stating that the integrated starlight transmitted by the coronagraph is smaller than $\epsilon$, but we have verified numerically that it yields a stellar throughput which is in most cases within 20\% of $\epsilon$. This can be explained by the very rapid decrease of $m_i$ with $i$ (especially for smaller stars) and the rapid quadratic increase of stellar leaks if $|c_i|$ goes above $\sqrt{\epsilon}/m_i$. For modest stellar sizes (below $\lambda/d$), equation \ref{equ:ciconst} constrains only the first few $c_i$ (the others are constrained by $|c_i|<1$). We note that adopting a zero stellar size will result in $m_0 = 1$ and $m_i = 0$ for $i>0$. In this case, equation \ref{equ:ciconst} only imposes a constraint for $i=0$. To obtain the maximum throughput as a function of angular distance, the vector $A(\alpha_p)$ is decomposed in the new base, and its first coefficients are multiplied by the values of $|c_i|$ derived in equation \ref{equ:ciconst}.
\begin{figure}[h]
\includegraphics[scale=0.33]{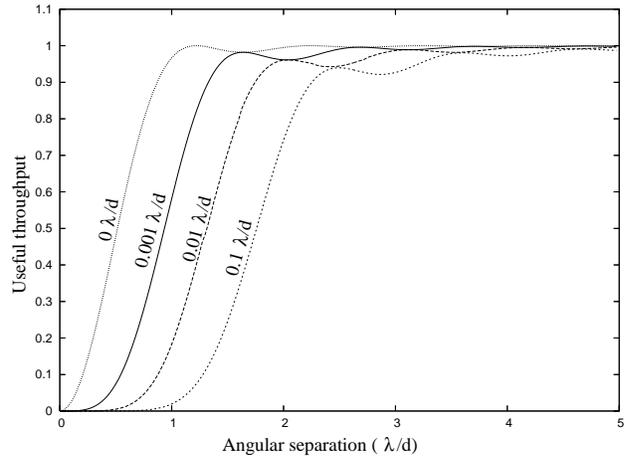}
\caption{\label{fig:upplimTT} Upper limit on the off-axis throughput of a coronagraph for different stellar radii.}
\end{figure}
The results obtained, shown in Figure \ref{fig:upplimTT} and also overplotted in Figure \ref{fig:usefulTtt}, confirm the large effect of a small increase in stellar size on the coronagraphic IWA. In the point source case, the 50\% throughput can be reached at 0.5$\lambda/d$; For a $0.1 \lambda/d$-radius source, the 50\% throughput limit moves to almost $2 \lambda/d$. This behavior is contrary to a ``geometrical optics'' intuition, which would suggest that increasing the stellar size by 0.1 $\lambda/d$ moves IWA by 0.1 $\lambda/d$. In the linear algebra model of coronagraphy presented in this work, this behavior is however understandable: as the stellar diameter increases, a higher number of modes needs to be removed from the wavefront; each mode removed increases the coronagraph's ``blind spot'' by a $\approx (\lambda/d)^2$ area in the focal plane.

\begin{figure*}[htb]
\includegraphics[scale=0.92]{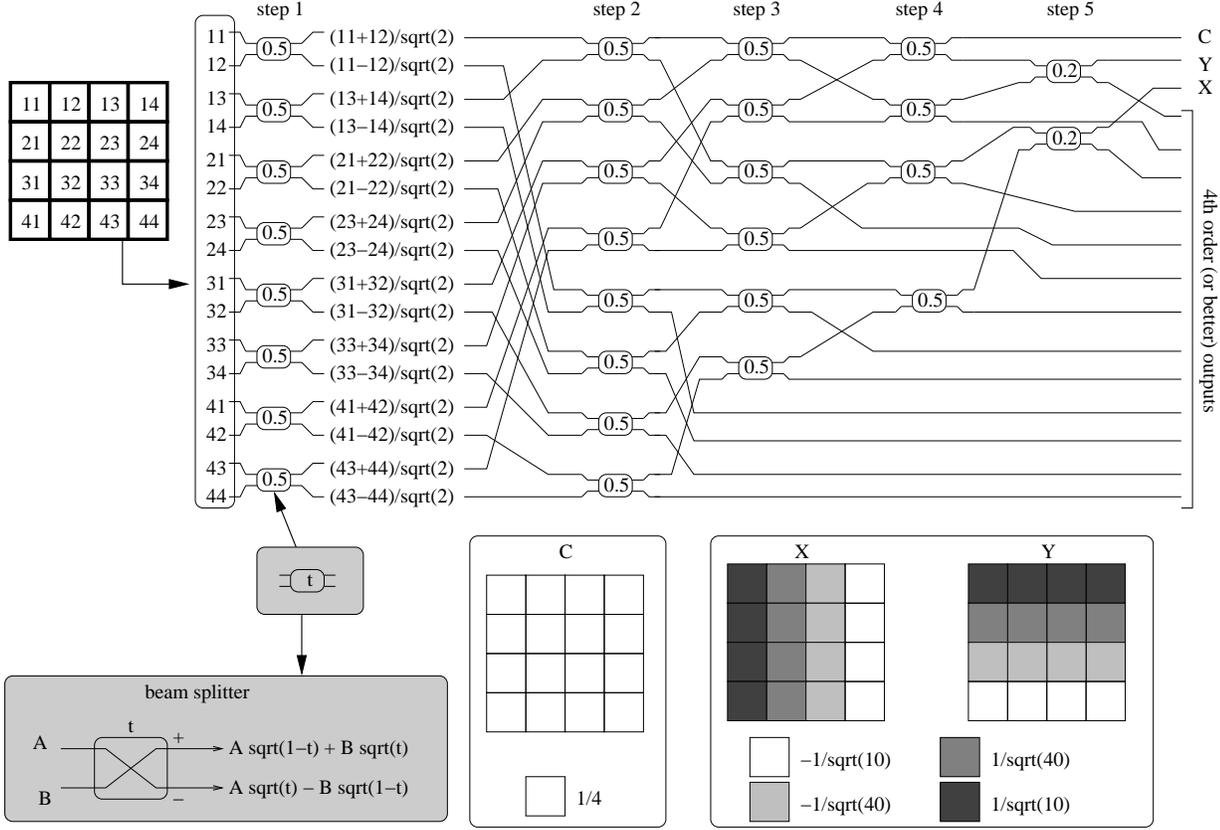}
\caption{\label{fig:dicc} Example of a beam splitter-based coronagraphs with $c_0 = c_1 = c_2 = 0$ (perfect rejection of the first 3 vectors $M_i$) designed for a square aperture. The telescope pupil (top left) is decomposed in a series of individual subpupils (shown in the input vector on the right of the pupil) which undergo interferometric combinations through beam splitters. The coronagraph outputs isolates the first 3 modes found in an extended source, as shown in the bottom right: C ($= M_0$), X ($= M_1$) and Y ($= M_2$). This coronagraph produces a $4^{th}$ order null and therefore provides some immunity to stellar angular size. The same technique can be generalized to circular pupil and better sensitivity to stellar angular size (more vectors $M_i$ isolated).}
\end{figure*}

A coronagraph which follows exactly the limit shown in Figure \ref{fig:upplimTT} can be assembled from discrete beam splitters and phase shifts as described in \S\ref{ssec:canitbebuilt}. An example of such a design is shown in Figure \ref{fig:dicc}. A $c_0=0$ coronagraph is often referred to as a $2^{nd}$ order null coronagraph, while $c_0=c_1=c_2=0$ insures a $4^{th}$ order null (and so on...). 

\subsection{Coronagraphic imaging at small angular separation ($< \lambda/D$)}

In this section, our linear algebra-based model is used to predict and explain how different coronagraph behave at small angular separation. These results allow us to understand how coronagraphic performance degrades as stellar angular diameter increases, and allow in \S\ref{ssec:specklenulling} to determine if wavefront control techniques can mitigate this problem.

\subsubsection{Second-order coronagraphs (IAC, PM, 4QPM)}
\label{sssec:2ndorderC}
\begin{figure}[htb]
\includegraphics[scale=0.68]{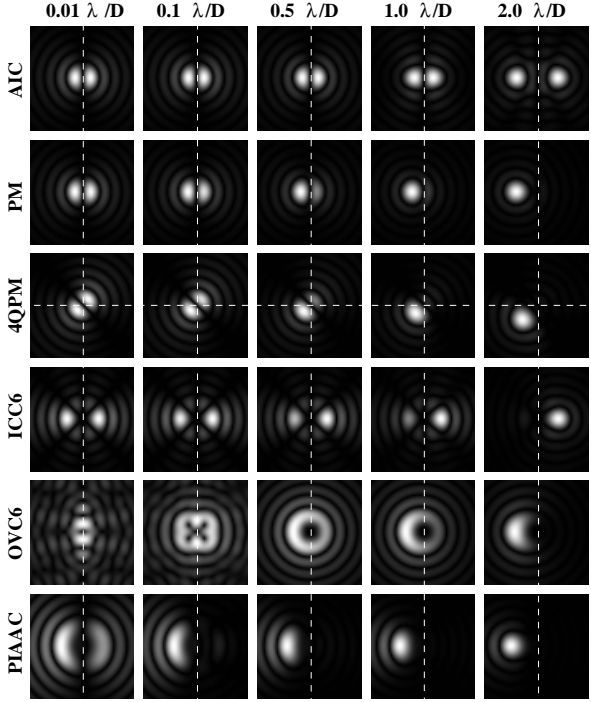}
\caption{\label{fig:PSFs_smallangle} PSFs of selected small-IWA coronagraphs as a function of angular separation. At small angular separation, the PSF of AIC, PM and 4QPM are centro-symmetric, making it impossible to determine on which side of the optical axis is the source. The OVC$_6$ PSFs at $0.01\lambda/D$ and $0.1 \lambda/D$ are dominated by numerical roundoff errors. Note that the brightness scale is not constant, as images on the left are much fainter than images on the right.}
\end{figure}
The smallest-IWA coronagraphs considered in this study (IAC, PM, 4QPM) all remove mode M0 (see Figure \ref{fig:modes}) from the incoming wavefront but ``transmit'' (at least partially) modes M1 and M2. At very small angular separations, the ``transmitted'' wavefront is therefore dominated by a combination of modes M1 and M2. The coefficients of this combination are proportional to the off axis distance, resulting in a coronagraphic residual light intensity proportional to the square of the source's angular separation ($2^{nd}$ order coronagraphs). It is therefore expected that for these coronagraphs, the complex amplitude in the final image is also proportional to the source coordinates: the complex amplitude for a source at position $(u,v)$ is the opposite of the complex amplitude for a source at position $(-u,-v)$. {\bf At small angular separations, our theoretical analysis therefore predicts that the images (square modulus of the complex amplitudes) of two point sources on opposite sides of the optical axis are identical, and that, as the source moves away from the optical axis, its image is simply multiplied by a constant, but does not change structure.} Both these predictions are confirmed by numerical simulations, as shown in Figure \ref{fig:PSFs_smallangle} (top 3 rows).

\subsubsection{Other small-IWA coronagraphs (ICC6, OVC6, PIAAC)}
Just as in the case of second-order coronagraphs, the linear algebra-based theory outlined in the previous sections can predict/explain the behavior of other coronagraphs at small angular separation. 

The Interferometric Combination Coronagraph (ICC6) which will be described in \S\ref{sec:idealDEF} perfectly removes the first 6 modes shown in Figure \ref{fig:modes}, but leaves other modes untouched. The ``transmitted'' wavefront of this theoretical coronagraph at small angular separation is therefore dominated by the $-i(xu+yv)^3/6$ term in the polynomial expansion of $e^{i(xu+yv)}$, corresponding to modes $x^3$, $y^3$, $xy^2$ and $x^2y$ (not shown in Figure \ref{fig:modes}). We can therefore predict that along any direction, its PSF is a fixed image multiplied by the 6$^{th}$ power of source angular separation, and that PSFs for 2 point sources on opposite sides of the optical axis will be identical.
A similar behavior is seen in Figure \ref{fig:PSFs_smallangle} for the OVC$_6$, which also removes the first 6 modes shown in Figure \ref{fig:modes}. Since this coronagraph also attenuates further modes, the PSF structures and throughput are however different from the ICC$_6$, which is designed to perfectly cancel the first 6 modes and fully transmit the others.

Finally, the PIAAC is a more complex (but also more representative of ``real world'' coronagraphs) case: it strongly attenuates, but does not fully remove, modes 1 to 6 of Figure \ref{fig:modes}. At very small angular separation (up to $\approx 10^{-3} \lambda/D$), mode $M0$ dominates the coronagraphic residual: the PSF is independent of source position. In the $10^{-3}$ to $10^{-2} \lambda/D$ range, a small residual in modes $M1$ and $M2$ dominates and the coronagraph behaves as a second order coronagraph (see \S\ref{sssec:2ndorderC} above). At a $0.1 \lambda/D$ separation and beyond, no single mode is dominating, and several partially transmitted modes interfere and/or are together to form the image: the PSF clearly shows on which side of the optical axis the source is, a behavior that cannot be achieved if the image is dominated by a single mode since individual modes are either symmetric or antisymmetric in complex amplitude (and therefore always symmetric in intensity). 

\subsection{Speckle nulling techniques, stellar size and pointing errors}
\label{ssec:specklenulling}
Active elements (Deformable Mirrors) can be used to create a coherent ``anti speckle halo'' which destructively interferes with the unwanted residual focal plane speckles. These speckle nulling techniques, first proposed by \cite{malb95}, were more recently studied by several authors \citep{codo04,labe04,guyo05b,bord06} and successfully tested in laboratory \citep{trau03}. 

\subsubsection{Can speckle nulling techniques mitigate coronagraphic leaks produced by stellar angular size ?}

The star can be modeled as a set of point sources uniformly distributed on the stellar disk (these sources are not mutually coherent). 
We first assume that a second order coronagraph is used, and we consider only one of these point sources (not at the center of the disk) which contributes to the coronagraphic leakage by adding an image on the focal plane. This particular image is fully coherent and can therefore be canceled by a speckle nulling technique. The shape/state of the active optical element(s) (DM for example) in the speckle control device will be such that the phase and amplitude of the ``anti speckle halo'' is the opposite (same amplitude, $\pi$ phase offset) of the leakage: destructive interference between the 2 will leave no light in the focal plane.
We now decide to ``freeze'' the state of the speckle nulling device and move the point source on the other side of the optical axis. This multiplies the complex amplitude of the coronagraphic leakage by $-1$ (\S\ref{sssec:2ndorderC}) but does not change the ``anti speckle halo'' (which is independent of source position within the small angular displacement considered). The leakage and ``anti speckle halo'' will therefore add rather than subtract because of a phase mismatch, and the leakage intensity will be 4 times higher than without speckle nulling.
To every point on the stellar disk corresponds an equally bright point on the other side of the optical axis: speckle nulling cannot reduce coronagraphic leaks in a second-order coronagraph. 

The same ``phase mismatch'' argument applies to $6^{th}$ order coronagraph such as $ICC_6$. Independently of this ``phase mismatch'' effect (which would not apply to a $4^{th}$ order coronagraph dominated by modes M3 to M5), coronagraphic leaks increase as the point on the stellar disk moves away from the optical axis, and there is therefore a similar ``amplitude mismatch'' effect when stellar points at different radii are considered.

The stellar angular size problem described in \S\ref{sec:ttsens} truly is a fundamental limit. 

\subsubsection{Pointing errors and low-order aberrations}
Pointing errors (or any higher order optics aberrations), on the other hand, produce a coherent leak which can be suppressed by wavefront control techniques. Low order aberrations can be measured from the coronagraphic leaks they produce. A more optimal strategy is to ``catch'' aberrations before they start to increase coronagraphic leaks by utilizing ``for free'' the large amount light that is blocked by the coronagraph. This can usually be done by re-imaging of the light reflected by the focal plane occulter (or, for phase mask coronagraphs, the light outside the geometric pupil).

We note that speckle sensing and nulling usually require a small spectral bandwidth, as the speckle coherence is low in white light: the observing spectral band may have to be split with dichroics in small ``bins'' prior to applying speckle nulling.

\section{Zodiacal and exozodiacal light}
\label{sec:zodi}

\subsection{Mixing of incoherent background light with the planet's light: the diffractive efficiency factor}
\label{sec:DEF}
The ``useful throughput'' introduced in \S\ref{sec:usefulT} accurately quantifies the coronagraphic performance in a system composed solely of a star and a single planet. A planetary system is however observed against incoherent zodiacal and exozodiacal light backgrounds. For simplicity, we assume here that both background components are constant across the field (this is an approximation for the exozodiacal light) and that they can be numerically subtracted to the photon noise level.

We consider a planet+background system (no central star) in which $\beta$ is the ratio between total planet light and total background light within the planet image (which we define here, for convenience, as the smallest focal plane area containing half of the planet's PSF light). We denote $\beta_0$ this value for a non-coronagraphic imaging telescope (Airy pattern). Coronagraphs unfortunately tend to decrease $\beta$ ($\beta < \beta_0$) by a factor which we denote the {\bf diffractive efficiency factor (DEF)}:
\begin{itemize}
\item{Pupil amplitude apodization and/or clipping in the Lyot plane increases the angular size of the planet image, therefore mixing more background light with the planet light.}
\item{The AIC delivers double images of the planet, therefore dividing $\beta$ by two.}
\item{Coronagraphs delivering PSFs in which a significant fraction of the light is outside the main diffraction core decrease $\beta$.}
\end{itemize}
Three factors determine the DEF: the width $w$ of the PSF core (more exactly, the square root of the ``useful'' PSF area), the fraction $f$ of the PSF energy in the diffraction core, and the total throughput $T$. The collectible planet energy is proportional to $fT$ while the background flux collected in the same focal plane area is proportional to $w^2T$: the DEF is therefore proportional to $f/w^2$.

The diffractive efficiency factors of most of the coronagraphs in Table \ref{tab:coronagraphs} are listed in Table \ref{tab:coronagraphs_eval}, along with other key characteristics. The DEF is especially low on the CPA and BL8 due to their poor angular resolution. As discussed in \S\ref{sec:CPAbinarymask}, a CPA with a binary mask (shaped pupil) could ``amplify'' background levels by another factor 2: the combined effect is then equivalent to dividing the planet/background ratio by $\approx 7$.

\subsection{Background level}
The zodiacal light background level is about $m_V=22.5\:arcsec^{-2}$ (can be less in some directions), and an Earth-like planet (6400km radius, 0.33 visible albedo, orbiting a Sun-like star at 1 AU), at maximum elongation, is $m_V = 24.1+5 \log(d)$ where $d$ is the system distance in $pc$. If we assume that the zodiacal, exozodiacal and planet spectra are identical, the distance at which, within the planet's image ($\lambda/d$ wide), zodiacal+exozodiacal background matches the planet's flux, is, for a favorable ``face-on'' geometry: 
\begin{equation}
d_c(pc) \approx \frac{2.3 \times DEF}{\sqrt{(1+2z)}} \frac{D(m)}{\lambda(\mu m)}
\end{equation}
where $z$ is the exozodiacal cloud dust content, normalized to our own system, and $DEF$ is the diffractive efficiency factor described in \S\ref{sec:DEF}. For example, with $z=3$, $DEF=0.5$, $\lambda=0.55\mu m$ and $D=4m$, $d_c = 4.5 pc$: for all but a few targets, photon noise from the background level will be the main source of noise, and will drive the exposure time required for successful detection of ETPs. At longer wavelength, the PSF size increases, therefore gathering more zodi+exozodi background and reducing $d_c$.

\begin{deluxetable}{ l c c c c c c c c c c c c c }
\tablewidth{0pt}
\tabletypesize{\footnotesize}
\tablecaption{\label{tab:coronagraphs_eval}Coronagraph Characteristics for $10^{10}$ contrast}
\tablehead{Coronagraph & \multicolumn{2}{c}{Throughput} & Angular & diffractive & IWA\tablenotemark{c} & \multicolumn{4}{c}{IWA(5\%) with stellar radius\tablenotemark{d}}\\ 
name & average & peak & resolution\tablenotemark{a} & eff. fact.\tablenotemark{b} & & 0 $\lambda/d$ & 0.001$\lambda/d$ & 0.01$\lambda/d$ & 0.1$\lambda/d$}
\startdata
AIC & 0.50 & 0.57 & 1.0 & 0.5 & 0.38 & 0.15 & 6.14 & $>8.0$ & $>8.0$\\
VNC/BL4(1) & 0.09 & 0.36 & 1.65\tablenotemark{e} & 0.37\tablenotemark{e} & 1.49 & 1.27 & 1.27 & 1.43 & $>8.0$\\
PSC & 0.18 & 0.73 & 1.0\tablenotemark{f} & 0.5\tablenotemark{f} & 1.13 & 0.79 & 0.79 & 3.24 & $>8.0$\\
CPA & 0.08 & 0.08 & 1.81 & 0.31\tablenotemark{g} & 4.20 & 4.42 & 4.42 & 4.42 & 4.47\\
PPA & 0.08\tablenotemark{h} & 0.33\tablenotemark{h} & 1.35 & 0.18 & 3.91 & 3.29 & 3.29 & 3.30 & 4.02\\
APLC & 0.20 & 0.20 & 1.22 & 0.67 & 2.41 & 2.01 & 2.17 & 4.02 & $>8.0$\\
APLC$_2$ & 0.27 & 0.27 & 1.11 & 0.81 & 1.64 & 1.26 & 1.78 & 4.71 & $>8.0$\\
VNC/BL4(2) & 0.35 & 0.70 & 1.19 & 0.71 & 3.02 & 2.17 & 2.20 & 2.21 & 3.77 \\
BL8 & 0.26 & 0.28 & 1.86 & 0.29 & 2.96 & 3.04 & 3.04 & 3.04 & 3.05\\
PM & 0.71 & 0.71 & 1.0 & 1.0 & 0.69 & 0.23 & 2.94 & 3.59 & $>8.0$\\
4QPM & 1.0 & 1.0 & 1.0 & 1.0 & 0.84 & 0.28 & 2.39 & $>8.0$ & $>8.0$\\
OVC$_2$ & 1.0 & 1.0 & 1.0 & 1.0 & 0.84 & 0.21 & 2.21 & $>8.0$ & $>8.0$\\
OVC$_4$ & 1.0 & 1.0 & 1.0 & 1.0 & 1.62 & 0.62 & 0.66 & 1.74 & $>8.0$\\
OVC$_6$ & 1.0 & 1.0 & 1.0 & 1.0 & 2.45 & 1.02 & 1.02 & 1.19 & 2.87\\
OVC$_8$ & 1.0 & 1.0 & 1.0 & 1.0 & 3.26 & 1.41 & 1.42 & 1.42 & 3.01\\
ODC & -\tablenotemark{i} & 0.70 & 1.18\tablenotemark{j} & 0.72\tablenotemark{j} & 0.97 & 0.50 & 4.59 & $>8.0$ & $>8.0$\\
PIAA & 1.0 & 1.0 & 1.0 & 1.0 & 1.88 & 1.28 & 1.38 & 1.66 & 2.0\\
\hline
{\bf ICC}$_6$ & {\bf 1.0} & {\bf 1.0} & {\bf 1.0} & {\bf 1.0} & {\bf 1.32} & {\bf 0.76} & {\bf 0.76} & {\bf 0.76} & {\bf 1.76}\\
\hline
\enddata
\tablenotetext{a}{Square root of the smallest focal plane area containing half of the planet's light. Normalized to 1.0 for the Airy pattern of an non-coronagraphic telescope. In the case of the AIC, the angular resolution is measured on one of the 2 images.}
\tablenotetext{b}{Diffractive efficiency factor measured at $20 \lambda/d$ unless otherwise specified.}
\tablenotetext{c}{Angular separation at which the planet's useful throughput first reaches half of the peak throughput. This is the standard definition of inner working angle (IWA): it assumes that the planet is favorably placed (no radial average) and that the star is point-like.}
\tablenotetext{d}{Angular separation at which the planet's radially averaged useful throughput first reaches 5\% for the stellar radius shown.}
\tablenotetext{e}{Measured at $2.36 \lambda/d$.}
\tablenotetext{f}{Measured at $1.77 \lambda/d$.}
\tablenotetext{g}{Higher if the apodization is performed by a binary mask}
\tablenotetext{h}{Only the main PSF diffraction peak is considered in the throughput}
\tablenotetext{i}{The outer part of the field is not transmitted in the ODC}
\tablenotetext{j}{Measured at $2 \lambda/d$.}

\end{deluxetable}

\subsection{Can an ``ideal'' coronagraph be built with $DEF=1$ ?}
\label{sec:idealDEF}
In \S\ref{sec:ttideal}, we have demonstrated that a coronagraph which delivers theoretically ideal throughput on a partially resolved star can be designed using beam splitters and phase shifts. The example shown in Figure \ref{fig:dicc} achieves this goal by confining starlight in a few interferometric outputs (as the required contrast and/or stellar diameter increases, the number of such outputs grows). However, no care was given to the distribution of planet light across the remaining outputs, and it is likely that background and planet light are highly mixed, resulting in high diffractive efficiency factor and poor imaging capabilities. We show here that a coronagraph can be designed to simultaneously deliver optimal throughput for a partially resolved star (see \S\ref{sec:ttideal}), full angular resolution imaging and $DEF=1$. 

In a classical non-coronagraphic imaging telescope, the system matrix U shown in Figure \ref{fig:matrixU} is a Fourier transform, offering full angular resolution imaging and $DEF=1$ but poor coronagraphic performance: we denote $U_i$ this ``imaging'' matrix. On the other hand, a ``coronagraphic'' matrix $U_c$ optimized for coronagraphic performance, such as the one shown in Figure \ref{fig:dicc}, efficiently isolates starlight but suffers from poor imaging capability and high diffractive efficiency factor.

The same iterative method used to produce the ``vectors'' C, X and Y in Figure \ref{fig:dicc} can be used to produce any predefined unitary matrix. Unitary matrices can also be cascaded: the input of a matrix, instead of being the telescope pupil, can be the output of another matrix. For example, it is possible to cascade a ``pupil to focal plane'' matrix $U_i$ with a ``focal plane to pupil'' $U_i^{-1}$ matrix, with no net result (the product of the two matrices is the unity matrix).

\begin{figure}[htb]
\includegraphics[scale=0.4]{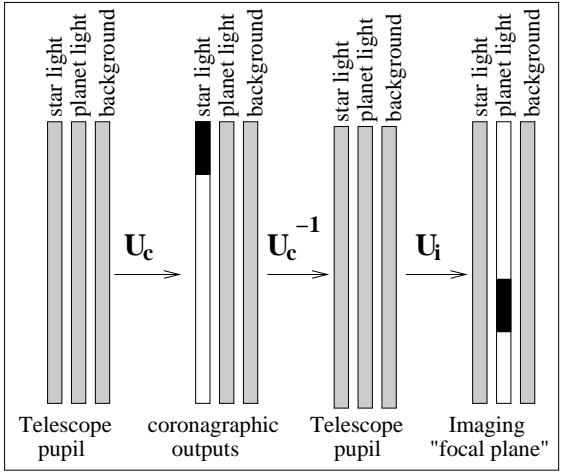}
\caption{\label{fig:dicc2} Design of an ``ideal'' coronagraph by cascading unitary matrices. In the first step, multiplication of the complex amplitude in the pupil plane by the coronagraphic matrix $U_c$ concentrates starlight in a few outputs. This operation can be inverted by multiplication by $U_c^{-1}$, therefore restoring the telescope pupil. Multiplication by $U_i$ produces a clean ``image'' of the planet. At each step, distribution of stellar (vector on the left), planet (center vector) and background (right vector) light is shown in greyscale: from white for no light to black for maximum light. The ``ideal'' coronagraph is obtained by simply masking the few outputs containing starlight in the ``coronagraphic outputs'' intermediate step.}
\end{figure}

One such cascade is shown in Figure \ref{fig:dicc2}: it is designed to produce a ``coronagraphic intermediate step'' where all starlight is concentrated in a few outputs and a final ``focal plane'' step with good imaging capability and $DEF=1$. The overall system matrix $U = U_c \times U_c^{-1} \times U_i = U_i$ is equivalent to a classical imaging telescope, as detection (conversion from amplitude to intensity) is performed after the last step of this chain. Stellar light can be easily masked in the ``coronagraphic intermediate step'', with minimal impact on the planet light (unless the planet is very close to the star) and background light (which is always spread across all outputs).
The sequence of optical steps to perform is therefore:
\begin{enumerate}
\item{Select the transformation $U_c$ which best isolates stellar light from planetary light}
\item{Block starlight in the few outputs where it has been concentrated.}
\item{Undo transformation 1 to recover the original telescope pupil (${U_c}^{-1}$)}
\item{Select the transformation which concentrates the planet light in a single bright diffraction peak (imaging $U_i$)}
\end{enumerate}
In an optimized implementation of this scheme, where the number of optical elements is to be minimized, there is no need to separate steps 2 and 3, as the intermediate product (the original telescope pupil) is not used.

An identical scheme is used in the PIAAC design, although this coronagraph falls short of the theoretically ideal performance derived in this work. The PIAAC first concentrates stellar light in a very tight diffraction core (step 1), which is then blocked by a mask (step 2). The pupil remapping used to perform step 1 unfortunately ``scrambles'' the planet image beyond $\approx 10 \lambda/d$. A second remapping unit is therefore then used to undo the first remapping (step 3). Finally, an image is formed (step 4).

Since the only currently known solution to build the ideal coronagraph requires many beam splitters, this type of coronagraph is referred to as the Interferometric Combination Coronagraph (ICC$_N$, where $N$ is the number of modes removed in the coronagraphic intermediate step) in this work. 

We have used the above method to simulate an ICC$_6$, designed to remove the 6 modes shown in Figure \ref{fig:modes}. The resulting coronagraphic throughput is of the $8^{th}$ order (just as for the BL8 coronagraph), with an IWA of 1.32 $\lambda/d$. This particular design is well suited for direct imaging of ETPs and accurately represents the absolute limit of what coronagraphy can achieve for this purpose. The useful throughput vs. angular separation of the ICC$_6$ is virtually identical to the ``theoretical limit'' shown in the upper panel (stellar radius $= 0.01 \lambda/d$) of Figure \ref{fig:usefulTtt}.

The performance of this ``ideal'' coronagraph (as well as the performance of a few other coronagraphs) is modeled in more detail in the following section. Table \ref{tab:coronagraphs_eval} concludes the previous sections by listing the main characteristics of the coronagraphs studied in this paper.

\section{Monte Carlo simulations}
\label{sec:montecarlo}

In the previous sections, coronagraph performance has been quantified by the {\bf useful throughput} and the {\bf diffractive efficiency factor}. Together, these two quantities could be used to evaluate, for a given target (star+planet+background), the signal-to-noise (SNR) as a function of exposure time. 
This simple model can unfortunately be quite inaccurate:
\begin{itemize}
\item{Exozodiacal light is not a smooth background (it is significantly brighter closer to the star and in a high inclination system, the exozodi disk can be a narrow linear feature): its effect cannot be accurately modeled by the diffractive efficiency factor. Planets are brighter when in nearly full phase, but are then closer to the star and therefore over brighter portions of the exozodi cloud.}
\item{In the presence of background light (zodiacal + exozodiacal light), faint parts of the PSFs that contribute to the useful throughput can have a negligible contribution to the detection SNR: the spatial distribution of the planet light affects the SNR.}
\end{itemize}
While the useful throughput and the DEF provided us with valuable insight into coronagraph performances, accurate estimation of the detection sensitivity requires explicit computation of images delivered by the coronagraph.

In this section, the performance of six promising coronagraphs identified in the previous sections (CPA, PIAAC, BL8, OVC6, ICC6 and VNC/BL4(2)) is quantified for the direct detection of Earth-type planets. Details of this simulation are given in \S\ref{sec:simumodel}. 
As the goal of this study is to show what can be potentially achieved with various coronagraphs, very optimistic assumptions have consistently been made. They are listed in Table \ref{tab:montecarlo} and further detailed in the next section.

\begin{deluxetable}{lccl}
\tabletypesize{\footnotesize}
\tablecaption{\label{tab:montecarlo} Simulation parameters.}
\tablehead{  &  value & unit &  notes}
\startdata
Optics throughput & 0.25 & & excludes losses due to coronagraph design\\
Wavefront quality & perfect & & no time-variable wavefront errors\\
Detector & perfect & & no readout noise or dark current\\
Imaging wavelength & 0.5 -- 0.6 & $\mu$m & \\ 
\\
Zodiacal background & 23.28 -- 22.24 & $m_V/arcsec^2$ & function of target ecliptic latitude\\
\\
1 zodi exozodiacal cloud brightness & 22.53 & $m_V/arcsec^2$ & at the habitable zone, for a face-on system\\
Exozodiacal cloud inner edge & 0.02$\sqrt{L}$ & AU & $L=$ star bolometric luminosity\\
Exozodiacal cloud outer edge & 5.0$\sqrt{L}$ & AU & $L=$ star bolometric luminosity\\
Exozodiacal cloud thickness & 0.0 & AU & thickness unresolved by telescope\\
exozodi optical depth & $\propto r^{-0.34}$ & & \\
\\
Planet radius & 6400 & km &\\
Planet albedo & 0.33 & & \\
semi-major axis & $\sqrt{L}$ & AU & $L=$ star bolometric luminosity\\
Eccentricity & 0 & & \\ 
Orbit inclination & $i=\pi/3$ & rad & statistical median for random orientation\\
\\
\enddata
\end{deluxetable}

\subsection{Description of the Simulation}
\label{sec:simumodel}

\subsubsection{Initial sample of targets}
\label{sec:targetsample}
\begin{figure}[htb]
\includegraphics[scale=0.3]{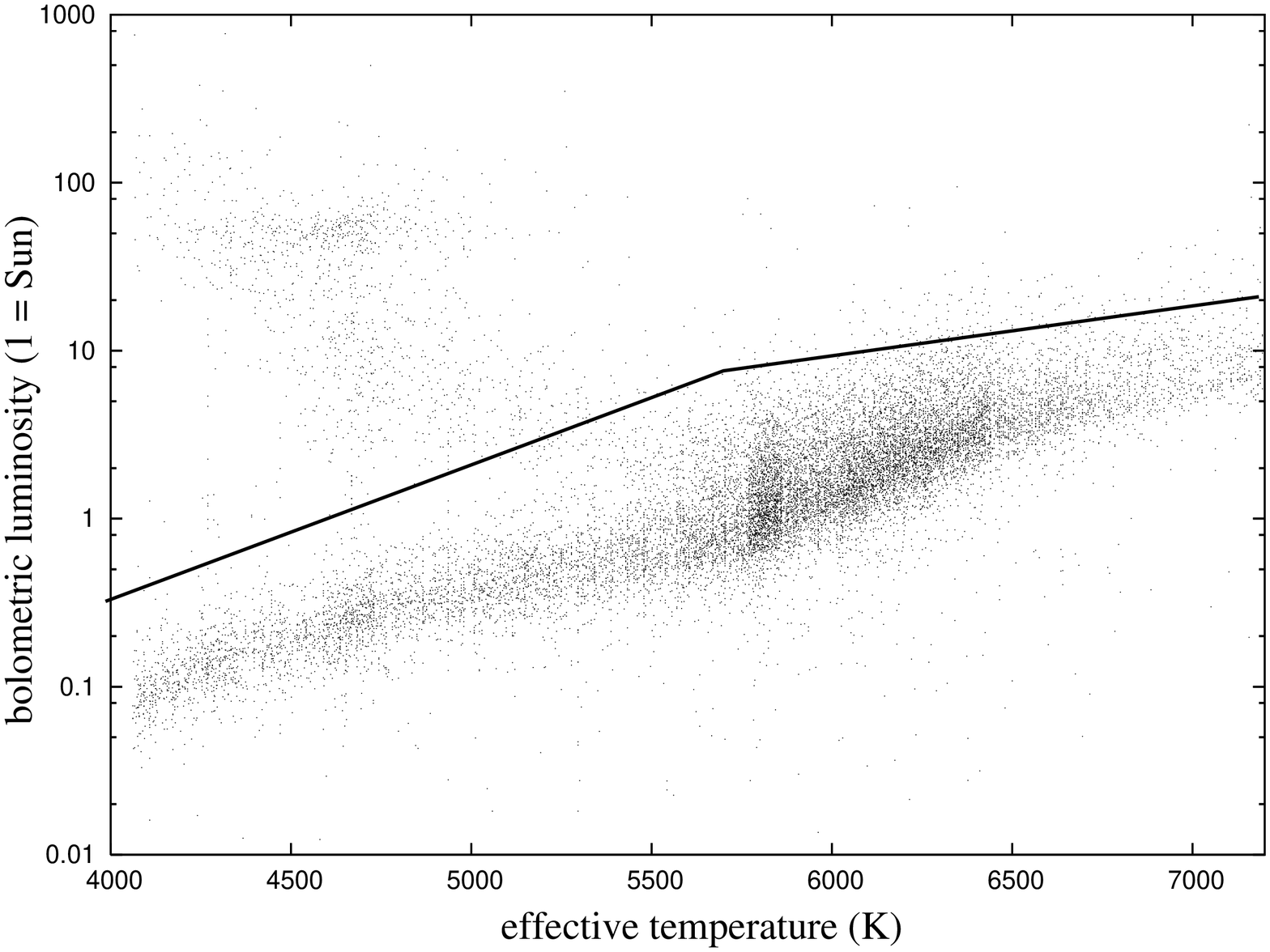}
\caption{\label{fig:lumTcr} Bolometric luminosity (1 = Sun) and effective temperature of stars within 100pc. The thick black line shows the criteria used in this work to exclude giants: all stars above this line are rejected from our sample of potential targets.}
\end{figure}

A list of potential target was first created by selecting stars within 100pc in the Hipparcos catalog. The effective temperature of each star was computed from the B-V color index, using the color index vs. temperature law of main sequence stars. Stars with effective temperature between 4060 K and 7200 K (types F, G and K) were then selected, and those listed as subarcsecond separation doubles removed. Giants were excluded from our sample using the stellar bolometric luminosity/effective temperature criteria outlined in Figure \ref{fig:lumTcr}.
This sample now contains 16921 stars. Each star is assumed to have an ETP. In this section, ETP (extrasolar terrestrial planet) is used to mean a standard Earth analog: Lambert phase function, radius = 6400km, albedo = 0.33, circular orbit at 1 AU times the square root of the star's bolometric luminosity (computed assuming the star is a perfect blackbody radiator). The inclination of the orbit, assumed to be coplanar with the exozodiacal disk, is set to $\pi/3$ for all systems: this is the median value expected for a random system orientation.

A final selection criteria is to reject systems for which the planet, at maximum elongation, is within 1 $\lambda/d$ of the star: none of the coronagraphs studied in this work can efficiently detect planets within this angular separation. For an 8m telescope observing in the 0.5--0.6 $\mu$m band, this final sample contains 3429 stars at an average distance of 51 pc.

\subsubsection{Zodiacal light}
Zodiacal light is modeled as a uniform background across the telescope's field of view. For each target, it is assumed that the observation occurs at the time of the year when the zodiacal background is minimal, which is about 40 days before or after opposition for most targets. The resulting zodiacal background, derived from \cite{leva80}, is a function of ecliptic latitude, and ranges from $m_V =$ 23.28 (ecliptic poles) to $m_V =$ 22.24 (ecliptic equator). 

The uniform zodiacal background is sampled by a square grid of points (0.2 $\lambda/d$ sampling in 2 orthogonal directions - about 30000 PSFs in the 20 $\lambda/d$ radius field of view). The PSF is computed for each point, and the final image delivered by the telescope/coronagraph is obtained by adding the ~30000 coronagraphic PSFs together.

\subsubsection{Exozodiacal light}
The exozodiacal cloud is assumed to be a thin disk with dust optical depth, as seen from above the 2D disk, varying as $r^{-0.34}$ (surface brightness $\propto r^{-2.34}$). This power law index is derived from modeling of our own zodiacal cloud in the Earth neighborhood \citep{kels98}. For a 1 zodi exozodiacal disk, the dust optical depth at the location of the planet is chosen to be identical to the vertical optical depth of our zodiacal cloud at 1 AU. In this model, the $0.55 \mu$m surface brightness of a face-on exozodi cloud at the location of the planet is thus a function of the ratio between the star bolometric luminosity (which defines where the planet is) and the star luminosity at $0.55 \mu$m (which defines how much the star illuminates the dust).
The inner edge of the exozodi cloud is set at 0.02 AU times the square root of the star's bolometric luminosity (dust sublimation), and its outer edge is 5 times the planet-star distance.

The image of the exozodiacal cloud delivered by the telescope/coronagraph is computed with the same 0.2 $\lambda/d$ sampling as used for the zodiacal background image.

\subsubsection{Detection metrics: Signal-to-Noise ratio (SNR) and detection probability}
It is assumed that the stellar diameter, exozodi cloud, zodiacal background and telescope/coronagraph characteristics are perfectly known. The planet detection SNR is therefore driven by photon noise and is proportional to the square root of the exposure time.
The photon noise (from the planet, star, zodiacal and exozodiacal images) and the signal (the planet image) are evaluated for each pixel of the image. The planet {\bf detection SNR} is obtained by the quadratic sum of the pixel SNRs: this optimal weighting scheme favors pixels with large planet flux and low zodi+exozodi+star flux.

We assume here that at the time of the observation, the position of the planet along its orbit is unknown. For a given exposure time, the planet detection SNR is computed at each point along the planet orbit. A limit on the SNR is adopted (SNR $>$ 7) for successful detection, and the number of points along the planet orbit for which this limit is reached yields the {\bf planet detection probability}. This probability, also referred to as completeness \citep{brow04,brow05}, is therefore a function of exposure time.  

A related metric used in this study, derived from the planet detection probability, is the single-exposure integration time required to reach a given planet detection probability.

\subsection{Example: HIP 56997}


\begin{figure*}[p]
\includegraphics[scale=0.7]{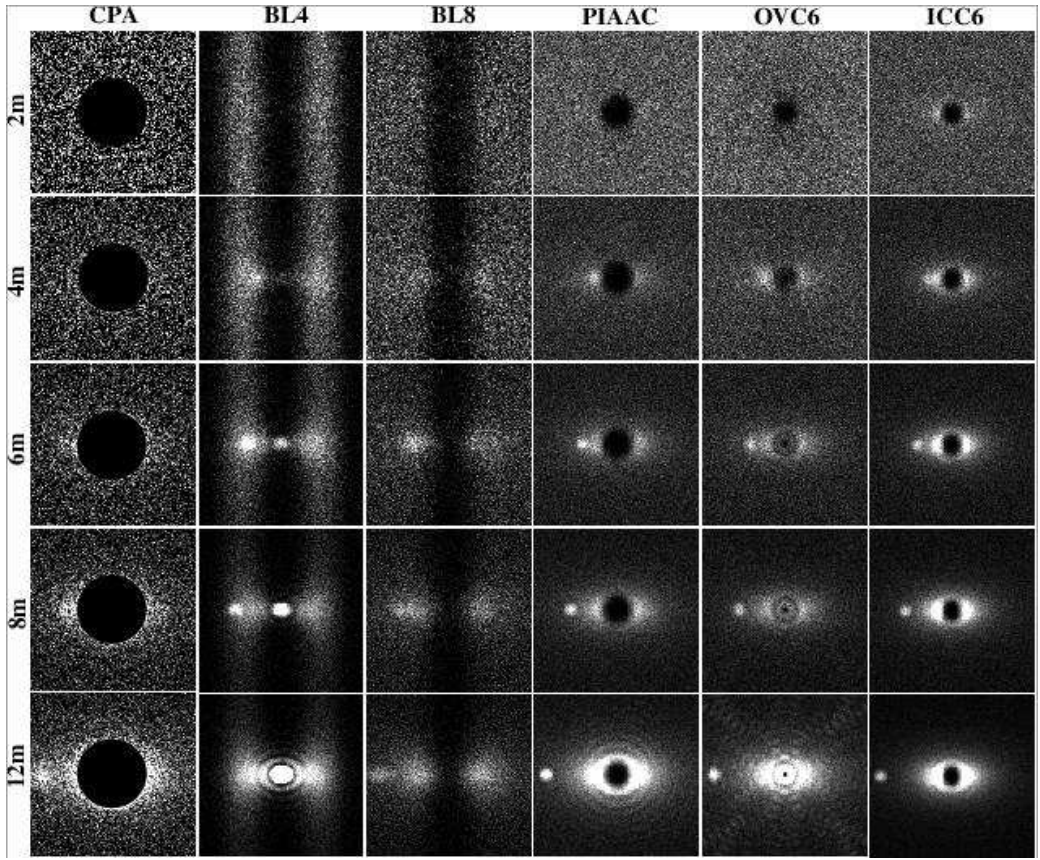}
\caption{\label{fig:exampleimages} Simulated 4 hours exposures of HIP 56997 and an hypothetical Earth-type planet at maximum elongation for telescope sizes ranging from 2m to 12m. HIP 56997 is a G8 type main sequence star at 9.54 pc. Each image assumes a perfect detector, minimum zodiacal background ($m_V =$ 22.95 zodiacal background for this 29 deg ecliptic latitude source), a 1 zodi exozodi cloud, a 25\% telescope+camera throughput, and a 0.1 $\mu$m bandpass centered at 0.55 $\mu$m. The system inclination for this particular simulation was arbitrarily set at $i \approx 59\:deg$. Each image is 20 x 20 $\lambda/d$, and the planet-star separation is 80 mas.}
\end{figure*}


Figure \ref{fig:exampleimages} shows some example frames obtained by the simulation code for the 6 coronagraphs selected. These images nicely illustrate coronagraphic characteristics quantified in the previous section:
\begin{itemize}
\item{{\bf Useful throughput at large separation.} The CPA, BL8, and to a lesser extent VNC/BL4(2) suffer from low coronagraphic throughput. As a result, the planet's image, even if well outside the coronagraph mask's influence, appears noisy (few photons detected). The PIAAC, OVC6 and ICC6, on the other hand, enjoy nearly 100\% throughput: the planet image is brighter and less noisy.}
\item{{\bf Angular resolution.} The CPA, BL8, and to a lesser extent VNC/BL4(2) have poorer angular resolution: the planet image is larger and more zodi/exozodi light is mixed with it.}
\item{{\bf Ability to work at small angular separation.} None of the coronagraph tested can detect the planet on a 2m telescope with the exposure time, wavelength, and throughput used in Figure \ref{fig:exampleimages}. Detection appears feasible on a 4m telescope with the VNC/BL4(2), PIAAC, OVC6 and ICC6, but requires a 6m telescope with the BL8. Finally, an 8m telescope is needed for detection with a CPA.}
\item{{\bf Sensitivity to stellar angular size.} The CPA and BL8 are extremely robust to stellar angular size: starlight leaks are virtually nonexistent even on the 12m telescope. With the VNC/BL4(2), PIAAC, OVC6 and ICC6, starlight is visible on the 12m telescope (equivalent to a 6m telescope observing the same system at 5pc), although it is still fainter that the exozodiacal contribution.}
\end{itemize}

Images similar to the ones shown in Figure \ref{fig:exampleimages} can be generated for each possible position of the planet along its orbit. These images have been used to compute for HIP 56997 the exposure time required to reach 20\% and 50\% detection (SNR=7) probability. Results, in Table \ref{tab:examplecase}, show very large discrepancies between coronagraph designs. For example, on a 6m telescope, the PIAAC requires 44 times less exposure time to reach a 50\% detection probability than the CPA. For this target, the performance of the PIAAC, OVC6 and ICC6 coronagraph are similar, while the other 3 coronagraph are significantly inferior (especially the CPA). For the VNC/BL4(2), it was assumed that the preferential direction of the coronagraph was optimally aligned with the system orientation.

\begin{deluxetable}{lcccccc}
\tablecaption{\label{tab:examplecase} Theoretical exposure times required to reach 20\% and 50\% detection (SNR=7) probability on HIP 56997 in the 0.5--0.6 $\mu$m band.}
\tablehead{\colhead{Tel. Diam.}  &  \colhead{CPA} & \colhead{VNC/BL4(2)} &  \colhead{BL8} & \colhead{PIAAC} & \colhead{OVC6} & \colhead{ICC6}}
\startdata
 & \multicolumn{6}{c}{{\bf 20\% detection probability}}\\
2m & \nodata\tablenotemark{a} & \nodata & \nodata & \nodata & \nodata & 12.4 hour\\
4m & \nodata & 2.1 hour & 11.4 hour & 34 min & 48 min & 14.3 min\\
6m & 6.9 hour & 14 min & 44.8 min & 5.7 min & 8.6 min & 3.78 min\\
8m & 63 min & 4.8 min & 17 min & 2.1 min & 3.1 min & 110 sec\\
12m & 12 min & 2.7 min & 4.3 min & 45 sec & 56 sec & 42 sec\\ 
\hline\\
 & \multicolumn{6}{c}{{\bf 50\% detection probability}}\\
2m    & \nodata & \nodata & \nodata & \nodata & \nodata & \nodata\\
4m    & \nodata & 12.7 hour & \nodata & 2.3 hour & 2.2 hour & 44.1 min\\
6m    & \nodata & 59 min & 5.0 hour & 14 min & 19.8 min & 12.23 min\\
8m    & 4.3 hour & 18 min & 1.8 hour & 5.8 min & 7.6 min & 5.4 min\\
12m    & 31 min & 55 min & 19 min & 2.0 min & 2.4 min & 113 sec\\

\enddata
\tablenotetext{a}{Exposure times above 1 day are not shown in this Table, as they are practically unrealistic.}
\end{deluxetable}

\subsection{4m and 8m diameter telescopes observing in the 0.5--0.6 micron band}
\begin{figure*}[p]
\includegraphics[scale=0.3]{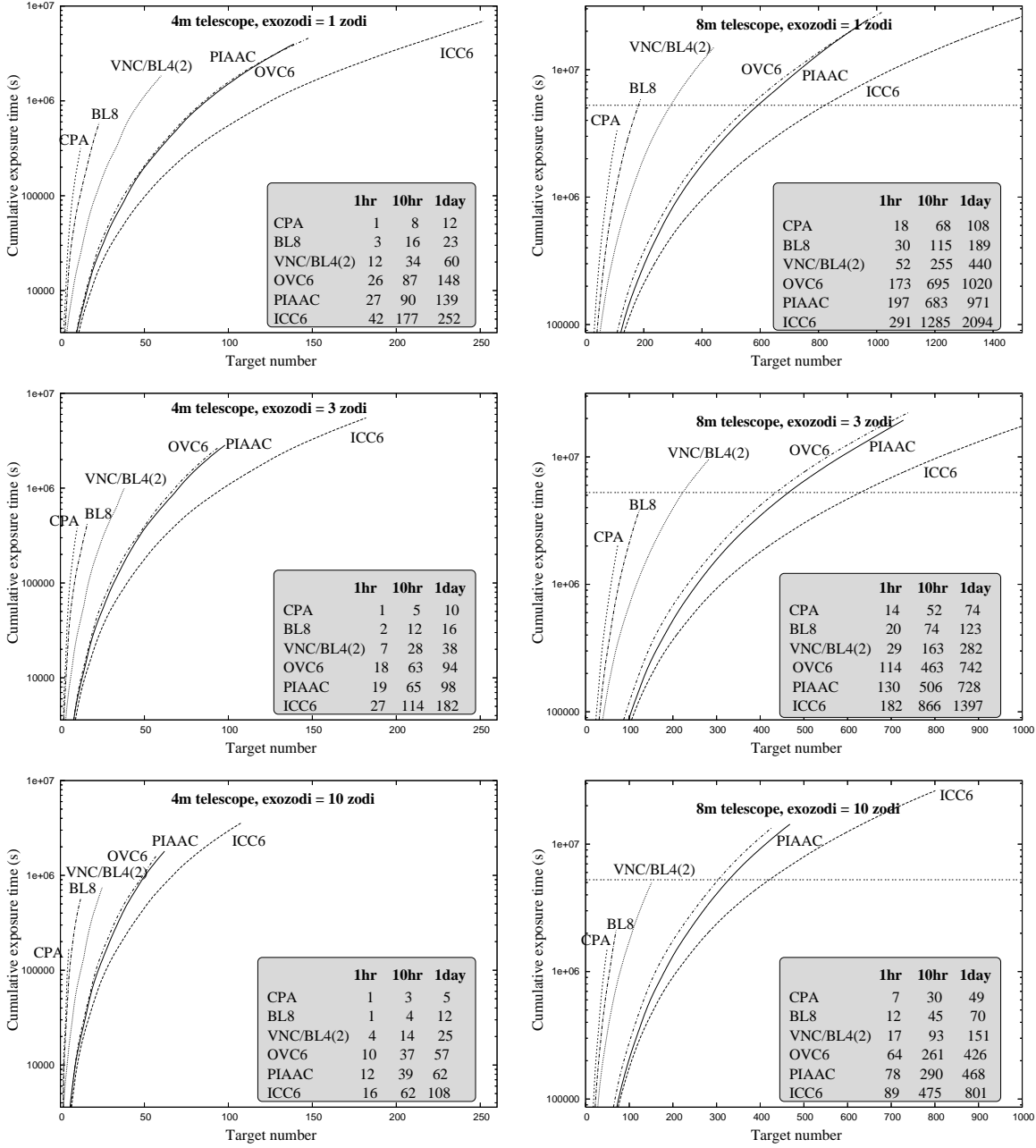}
\caption{\label{fig:idealcaseetime} Total cumulative exposure times required to reach a 50\% planet detection (SNR=7) probability for a single observation as a function of number of targets. For all simulations, a 25\% throughput in the 0.5$\mu$m--0.6$\mu$m band is adopted, and targets are ordered with increasing exposure time. Each curve terminates when the required exposure time per target reaches 1 day. Results are shown for a 4m telescope (left) and an 8m telescope (right) with exozodi levels ranging from 1 zodi (top) to 10 zodi (bottom). The number of accessible targets for which the required exposure times are less than 1 hour, 10 hour and 1 day are listed for each case in the Grey boxes. In the 8m telescope plots, the horizontal line corresponds to a 2 months ``shutter open'' cumulative exposure time, and may be considered as a practical limit on the number of targets that can be visited.}
\end{figure*}

In the previous section, a single target was adopted while the telescope size was variable. Here, we adopt fixed telescope sizes and evaluate the telescope/coronagraph combination performance on a sample of nearby stars, defined in \ref{sec:targetsample}. Results are shown in Figure \ref{fig:idealcaseetime} for 4m and 8m telescopes.

In a purely background-limited detection, exposure time is expected to scale linearly with the background level: approximately 7 times longer for the 10 zodi exozodi case than for the 1 zodi exozodi case (assuming the local zodi background corresponds to half a zodi of exozodiacal light). The fact that this is generally true in Figure \ref{fig:idealcaseetime} confirms the importance of the DEF in coronagraph designs. For coronagraphs with low DEFs, such as PIAAC, OVC6 and ICC6, increasing the exozodi however results in moderate increase of exposure time for nearby targets (non background-limited detections). This effect is most clearly visible in the 8m telescope case for the easiest targets.

Figure \ref{fig:idealcaseetime} compares how many targets fall below the 1hr, 10hr and 1 day required exposure time levels with different exozodi contents. With the 4m telescope, the CPA and BL8 cannot access the ``habitable zone'' of targets beyond $\approx 5 pc$, resulting in a very small number of potential targets (8 and 16 targets respectively for CPA and BL8 if ``potential target'' is defined as one that requires less than 10hr exposure time to reach the 50\% detection probability). The VNC/BL4(2), thanks to its higher throughput, can access 34 targets. With this telescope diameter (4m), the stellar angular size is sufficiently small to allow improved performance over CPA and BL8, which are more robust to stellar angular size. In increasing order of performance, the OVC6, PIAAC and ICC6 offer from 87 to 177 targets in the 1-zodi case. It is interesting to note that the performance of the OVC6 and PIAAC is very similar, while the ``ideal'' ICC6 offers almost twice as many targets as the PIAAC. This is mostly due to the benefit of smaller IWA in the ICC6 over the PIAAC: $1.32 \lambda/d$ vs. $1.88 \lambda/d$, which increases the IWA-accessible volume by 2.9. The planet intrinsic brightness and the background level prevent access to distant systems, and there is therefore little gain in reducing the IWA below $1.88 \lambda/d$.

With any of the 3 ``top'' coronagraph designs simulated in this work (PIAAC, OVC6 and ICC6), a 4m telescope could, with a 10hr per observation limit and a 1-zodi assumption on the exozodi, perform a $\approx 100$ target survey for ETPs. The total exposure time to complete this survey, assuming 6 observations of 10hr for each target, would be well under a year, and therefore sounds realistically achievable. Under the same assumptions, strong exozodi (10 zodi) would cut in half the sample size.

Figure \ref{fig:idealcaseetime} shows that doubling the telescope diameter (from 4m to 8m) multiplies the number of accessible targets by about 8 for all exozodi content, coronagraph type and exposure time limit combinations tested. The relative efficiencies of the 6 coronagraphs tested stay the same, with, in order of decreasing performance, the ICC6, PIAAC and OVC6 able to access several hundred targets in the 1-zodi case (with a 1hr exposure time limit criteria), the VNC/BL4(2) about 50 targets and the BL8 and CPA a few tens of targets.

In this study, we have used an Earth-analog as a prototype for Earth-like planets.  The broad concept of candidate Earth-like (potentially habitable) planets likely extends to include planet diameters to several times smaller and larger than Earth.  Orienting a survey to optimize detection for a different planet radius would of course change the observation times for each system and the number of systems which could be surveyed. Similarly, an observing program can be optimized for any target list and planet description to maximize the productivity with respect to total detections or other criterion.

\section{Conclusion}

We have quantified the performance limit imposed by fundamental physics on the direct imaging of ETPs. A solution to build an ``optimal'' coronagraph reaching this limit was proposed, but may be technologically difficult to implement. Fortunately, we have also demonstrated that two existing coronagraph designs (PIAAC and OVC6) yield performances reasonably close (within a factor 2 in number of accessible targets) to this fundamental limit. These recently proposed designs could image ETPs around up to about 100 stars with a 4m telescope, or several hundred stars with an 8m telescope.

For the coronagraphs to perform as detailed in this study, two serious technological challenges must however be solved:
\begin{itemize}
\item{{\bf Wavefront control at the sub-\AA$\:$level}. Thanks to the large number of stellar photons available, recent detector developments, and the expectation that a space telescope can be designed to offer a relatively stable wavefront (with little rapidly varying aberrations, as opposed to atmospheric turbulence for ground-based telescopes), it may be reasonable to hope for this level of correction. Several authors have recently explored promising wavefront control schemes free of aliasing and non-common path aberrations \citep{codo04,labe04,guyo05b,bord06}, and contrast levels obtained in laboratory \citep{trau03} are approaching the desired performance.}
\item{{\bf Chromatic effects} can seriously limit the coronagraph's usable spectral band. A large number of solutions have been proposed to ``achromatize'' coronagraphs, many of them specific to a single coronagraph design \citep{abe01,soum03a,aime05a,mawe05,pluz06,swar06}. In addition to coronagraph-induced chromatic effects, the beam delivered by the telescope is expected to be chromatic in amplitude and optical pathlength differences (non-ideal coatings, chromatic diffraction propagation between optics). In general, chromatic effects can be mitigated by reducing chromaticity in individual components (beam splitters, phase-shifters, apodization masks) or/and by splitting the spectral band into several narrower bands with dichroics.}
\end{itemize}

More fundamentally, the assumption that detection is photon noise limited seems highly optimistic: ETPs are embedded in an exozodiacal light unlikely to be perfectly smooth and symmetrical. Since good angular resolution may help to distinguish between a planet and exozodiacal light features (such as arcs or rings), it is important to avoid using coronagraphs that clip or strongly attenuate the edge of the telescope pupil if at all possible.

The results obtained in this work should therefore be cautiously considered as upper limits on what may be achieved with a particular coronagraph/space telescope design for the specific goal of imaging ETPs around nearby stars. We also note that several low throughput/large IWA coronagraphs studied in this paper can be more favorably designed for ground-based applications, where contrast requirements are relaxed by about 4 orders of magnitude.  

The linear algebra approach to coronagraphy presented in this work successfully explains key characteristics and fundamental limits of coronagraphy. While we have focused our study on the fundamental limitations of coronagraphy (due to stellar angular size and background contamination), this model can be extended to evaluate the sensitivity of coronagraphs to wavefront aberrations other than tip-tilt. The same model can also be used to derive fundamental limits and guide the design of coronagraphs for more ``exotic'' pupil shapes, from segmented pupils telescopes to short or long baseline interferometric arrays.

\acknowledgements
This work was carried out with the support of the National Astronomical Observatory of Japan and JPL contract numbers 1254445 and 1257767 for Development of Technologies for the Terrestrial Planet Finder Mission. We thank Lyu Abe and Remi Soummer for providing comments and helping the authors to understand tradeoffs for the Apodized Pupil Lyot Coronagraph. We thank Stuart Shaklan for his useful comments and review of this work.

\end{document}